\documentclass[10pt,final,conference]{IEEEtran}
\IEEEoverridecommandlockouts
\usepackage{cite}
\usepackage{amsmath,amssymb,amsfonts}
\usepackage[ruled,linesnumbered]{algorithm2e}
\usepackage{graphicx}
\usepackage{textcomp}
\usepackage{multirow}
\usepackage{xcolor}
\def\BibTeX{{\rm B\kern-.05em{\sc i\kern-.025em b}\kern-.08em
    T\kern-.1667em\lower.7ex\hbox{E}\kern-.125emX}}

\usepackage{hyperref}
\usepackage{xspace}
\usepackage{booktabs}
\usepackage{enumitem}
\usepackage[font=footnotesize]{caption}
\usepackage[font=footnotesize]{subcaption}
\usepackage{url}
\usepackage{forest}

\usepackage{xcolor}
\makeatletter
\newcommand{\timepeak}{{\small \textsf{time}_{\textsf{peak}}}\xspace}
\newcommand{\timestart}{{\small \textsf{time}_{\textsf{start}}}\xspace}
\newcommand{\proj}{\textsc{Lightor}\xspace}
\newcommand{\dota}{{\small \textsf{Dota2}}\xspace}
\newcommand{\toretter}{{\small \textsf{Toretter}}\xspace}
\newcommand{\socialskip}{{\small \textsf{SocialSkip}}\xspace}
\newcommand{\moocer}{{\small \textsf{Moocer}}\xspace}
\newcommand{\chatlstm}{{\small \textsf{Chat-LSTM}}\xspace}
\newcommand{\jointlstm}{{\small \textsf{Joint-LSTM}}\xspace}
\newcommand{\precisionK}{{\small \textsf{Precision@K}}\xspace}
\newcommand{\chatprecisionK}{{\small \textsf{Chat Precision@K}}\xspace}
\newcommand{\chatprecisionten}{{\small \textsf{Chat Precision@10}}\xspace}

\newcommand{\videoprecisionstartK}{{\small \textsf{Video Precision@K (start)}}\xspace}
\newcommand{\videoprecisionendK}{{\small \textsf{Video Precision@K (end)}}\xspace}
\newcommand{\lol}{{\small \textsf{LoL}}\xspace}
\newcommand{\watch}{{\small \textsf{play}}\xspace}

\newcommand{\Rmnum}[1]{\uppercase\expandafter{\romannumeral #1\relax}}
\newcommand{\textsfsmall}[1]{\textsf{\small #1}}
\makeatletter

\begin{document}

\title{Towards Extracting Highlights From Recorded Live Videos: An Implicit Crowdsourcing Approach
}


\author{
\IEEEauthorblockN{Ruochen Jiang\IEEEauthorrefmark{1}\IEEEauthorrefmark{3}}
\IEEEauthorblockA{
\textit{\IEEEauthorrefmark{2}Simon Fraser University}\\
\{changboq, jnwang\}@sfu.ca}

\and
\IEEEauthorblockN{Changbo Qu\IEEEauthorrefmark{1}\IEEEauthorrefmark{2} \hspace{3em} Jiannan Wang\IEEEauthorrefmark{2}}
\thanks{\IEEEauthorrefmark{1} Work done at SFU. Both authors contributed equally to this research.}
\IEEEauthorblockA{
\textit{\IEEEauthorrefmark{3}Ohio State University}\\
jiang.2091@osu.edu}

\and
\IEEEauthorblockN{Chi Wang}
\IEEEauthorblockA{\textit{Microsoft}\\
wang\_chi@microsoft.com}
\and
\IEEEauthorblockN{Yudian Zheng}
\IEEEauthorblockA{\textit{Twitter}\\
yudianz@twitter.com}
}

\maketitle

\begin{abstract}
Live streaming platforms need to store a lot of recorded live videos on a daily basis. An important problem is how to automatically extract highlights (i.e., attractive short video clips) from these massive, long recorded live videos. One approach is to directly apply a highlight extraction algorithm to video content. However, algorithmic approaches are either domain-specific, which require experts to spend a long time to design, or resource-intensive, which require a lot of training data and/or computing resources. In this paper, we propose \proj, a novel implicit crowdsourcing approach to overcome these limitations. The key insight is to collect users' natural interactions with a live streaming platform, and then leverage them to detect highlights. \proj consists of two major components. Highlight Initializer collects time-stamped chat messages from a live video and then uses them to predict approximate highlight positions. Highlight Extractor keeps track of how users interact with these approximate highlight positions and then refines these positions iteratively. We find that the collected user chat and interaction data are very noisy, and propose effective techniques to deal with noise. \proj can be easily deployed into existing live streaming platforms, or be implemented as a web browser extension. We recruit hundreds of users from Amazon Mechanical Turk, and evaluate the performance of \proj using two popular games in Twitch. The results show that \proj can achieve high extraction precision with a small set of training data and low computing resources.
\end{abstract}


\section{Introduction}
Video data is booming and will account for 90\% of all internet traffic by 2020 as predicted by Cisco~\cite{index2016cisco}. Improving video-related services is of growing interest in the database and data mining community~\cite{Wu:2014:CTV:2623330.2623625, lee2018collaborative,DBLP:journals/pvldb/KangEABZ17,krishnan2018deeplens,DBLP:journals/pvldb/HaynesMABCC18,DBLP:conf/sigmod/LuCKC18}. As an important type of video service, live streaming platforms such as Twitch, Mixer, YouTube Live, and Facebook Live fulfill the mission of democratizing live video broadcasting. 
With these platforms, anyone can be a broadcaster to record a video and broadcast it in real time; anyone can be a viewer to watch a live video and chat about it in real time. This unique experience makes these platforms more and more popular nowadays. For example,  by 2018, Twitch has reached 3.1 million unique monthly broadcasters, and over 44 billion minutes of videos are watched each month~\cite{twitch-stats}.

Once a live stream is complete, the recorded video along with time-stamped chat messages will be archived. A recorded video is often very long (from half an hour to several hours). Many users do not have the patience to watch the entire recorded video but only look for a few \emph{highlights} to watch. A highlight represents a small part of the video that makes people feel excited or interested, and it typically lasts from a few seconds to less than one minute.  For example, a highlight in a \dota game video could be an exciting battle or a critical knockdown.

We study how to automatically extract highlights from a recorded live video. Finding a good solution to this problem could have a propound impact on live-streaming business. First, since it save users' time in manually finding the highlights, more and more users might be willing to watch recorded live videos, thus increasing the user engagement of a live-streaming platform. Second, highlight extraction is a fundamental task in video processing. If a live-streaming platform knows the highlights of each video, it will significantly improve the user experience of other profitable video applications, such as video search and video recommendation. 

One direction to solve this problem is to adopt a machine-only  approach~\cite{DBLP:conf/mm/RuiGA00,DBLP:journals/tip/EkinTM03,DBLP:conf/iccv/YangWLWGG15,DBLP:conf/cvpr/YaoMR16,DBLP:conf/emnlp/FuLBB17}. However, many existing highlight-detection algorithms~\cite{DBLP:conf/mm/RuiGA00,DBLP:journals/tip/EkinTM03} are domain specific. For example, to apply them to Twitch, domain experts need to design a different algorithm for every game type (e.g. \dota, \lol, etc.), which is not easy to develop and maintain. Some recent papers study how to train a deep-learning model (e.g., CNN, RNN, LSTM) to detect highlights~\cite{DBLP:conf/iccv/YangWLWGG15,DBLP:conf/cvpr/YaoMR16,DBLP:conf/emnlp/FuLBB17}. However, for each game type, they need to label a large number of videos in order to train a good model. The model trained on one type of game (e.g., \dota) usually does not generalize well to another (e.g., \lol). Furthermore, the training process is computationally expensive, which take several days to train a model using expensive GPUs. 


\begin{figure*} \vspace{-2.5em}
    \centering
    \includegraphics[scale = 0.47]{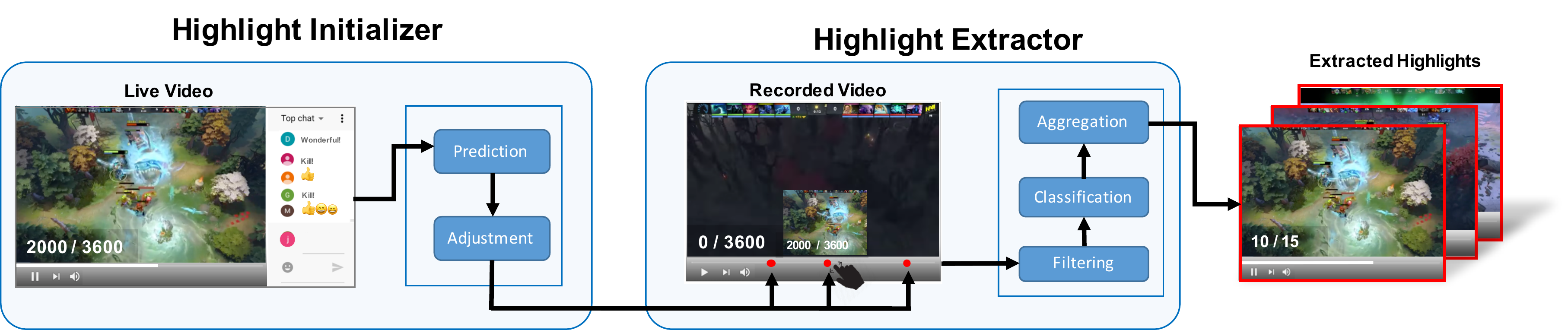}
    \caption{\proj: An implicit crowdsourcing workflow for extracting highlights from a recorded live video.}
    \label{fig:Overview}\vspace{-1.5em}
\end{figure*}

Crowdsourcing is a potential approach to overcome these limitations. Crowdsourcing seeks to recruit a large number of Internet workers to solve challenging problems (e.g., entity resolution~\cite{DBLP:journals/pvldb/WangKFF12,DBLP:journals/pvldb/WhangLG13}, image labeling~\cite{DBLP:conf/chi/AhnD04}). A simple solution is to adopt \emph{explicit} crowdsourcing, which pays workers some money and asks them to extract highlights from videos. Since this solution uses monetary incentives to recruit workers, it does not scale well to a large number of videos in live streaming platforms. 

In this paper, we propose a novel \emph{implicit crowdsourcing} to tackle this problem~\cite{implicit-crowdsourcing}. Implicit crowdsourcing is the idea of collecting implicit feedback from users (i.e., user's \emph{natural} interactions with the system) and then leveraging the feedback to solve a challenging problem. It has achieved great success in many domains. For example, reCAPTCHA~\cite{von2008recaptcha} leverages this idea to digitize old books. Search engines collect implicit clickthrough data to optimize web search ranking~\cite{DBLP:conf/kdd/Joachims02}. To apply this idea, we face two challenges. The first one is how to design an implicit crowdsourcing workflow so that video viewers interact with the system naturally but provide useful feedback implicitly. The second one is how to use the implicit (and noisy) feedback to detect and extract video highlights.  We address these two challenges as follows.

\vspace{.5em}

{ \noindent {\bf Implicit Crowdsourcing Workflow.}} We design a novel implicit crowdsourcing workflow, called  \proj. \proj consists of two components. i) Highlight Initializer takes a recorded live video as input and uses its time-stamped chat messages to detect which part of the video could have a highlight. For example, when a large number of chat messages pop up within a short period of time, users may talk about a highlight that has just happened. Note that Highlight Initializer can only get an approximate position of a highlight. It is still not clear about the exact boundary (i.e., exact start and end points) of a highlight. ii) Highlight Extractor is designed to address this problem. At each approximate position, it puts a ``red dot'' on the progress bar of the video, which informs users that there could be a highlight at this position. Note that users will not be forced to watch this highlight. Instead, they can watch the video as usual. Highlight Extractor collects user interaction data w.r.t. each red dot to identify the exact boundary of each highlight. 

\vspace{.5em}

{\noindent  {\bf Noisy User Data.}} 
One major challenge in our implicit crowdsourcing design is how to handle the high noise in the implicit feedback from users. For example, in Highlight Initializer, when a user leaves a chat message, she might not comment on the video content but chat with other users. In Highlight Extractor, when a user watches a certain part of video, she might not be attracted by the video content but check whether this part of video has something interesting. Therefore, we have to be able to separate noise from signal. We analyze real-world user data and derive a number of interesting observations. Based on these observations, we develop several effective techniques to deal with  noisy user interaction data.

\vspace{.5em}

\sloppy

\proj can be easily deployed on an existing live streaming platform. The only change is to add red dots to the progress bar of recorded videos. Based on a user survey, we find that most users prefer this change since red dots help them find more interesting highlights.
Furthermore, \proj can be implemented as a web browser extension, which has the potential to support any platform. 

We recruit about 500 game fans from Amazon Mechanical Turk and evaluate \proj using two popular games (\dota and \lol) from Twitch. The results show that (1) our proposed techniques make \proj achieve very high precision (up to $70\%-90\%$) in the returned top-$k$ highlights, which changes the system from unusable to usable, and (2) \proj requires 123$\times$ fewer training examples and over 100000$\times$ less training time compared to the state-of-the-art deep learning based approach, thus it is much preferable when there is a lack of training data or computing resources.

To summarize, our contributions are:

\begin{itemize}[leftmargin=*]\setlength\itemsep{0.25em}
    \item We study how to leverage implicit crowdsourcing to extract highlights from a recorded live video. We propose \proj, a novel workflow to achieve this goal. 
    \item We analyze real-world chat messages, and derive a number of interesting observations. Based on these observations, we develop a simple but effective Highlight Initializer.
    \item We derive a number of interesting observations from real-world user interaction data, and propose a novel Highlight Extractor to identify the exact boundary of each highlight. 
    \item  We discuss how to deploy \proj on an existing live streaming platform or implement it as a web extension. We recruit hundreds of users and evaluate \proj using real live video data.  The results demonstrate the superiority of \proj over baseline approaches. 
\end{itemize}

The remainder of this paper is organized as follows.  Section~\ref{sec:rw} reviews the related work. Section~\ref{sec:workflow} presents the \proj workflow. We discuss how Highlight Initializer and Highlight Extractor are built in Section~\ref{sec:init} and Section~\ref{sec:extract}, respectively.  Section~\ref{sec:deploy} discusses how to deploy \proj in practice. Experimental results are presented in Section~\ref{sec:exp}. We discuss our findings and lessons learned in Section~\ref{sec:lessons}, and present conclusions and future work in Section~\ref{sec:conc}. We provide a reproducibility report and release all the code and datasets at the project page: \textsf{\small \url{http://tiny.cc/lightor}}.

\section{Related Work}
\label{sec:rw}

Crowdsourcing has been extensively studied in the database community in recent years. However, prior work is mainly focused on explicit crowdsourcing~\cite{DBLP:journals/tkde/LiWZF16,DBLP:conf/sigmod/TongWZCDY18,DBLP:conf/sigmod/AmsterdamerDMNS14} or non-video-tasks~\cite{DBLP:conf/icde/HuLBCF16, DBLP:journals/pvldb/TongCS17, pournajaf2014spatial, hu2016crowdsourced,novgorodov2019generating}. This paper shows that exploring the use of implicit crowdsourcing for video tasks is a promising direction, calling for more attention to this exciting research direction.  In fact, our work touches a wide range of research topics in other communities. 
\vspace{.25em}

{\noindent \bf Computer Vision.} There is a recent trend to apply deep learning to highlight detection~\cite{DBLP:journals/corr/Song16a,DBLP:conf/cvpr/YaoMR16,DBLP:conf/emnlp/FuLBB17}. For example, a frame-based CNN model~\cite{DBLP:journals/corr/Song16a} was trained to detect the frames with significant visual effects for e-sports. In \cite{DBLP:conf/emnlp/FuLBB17}, a joint model of CNN on video and LSTM on chat was trained to detect highlights in game videos. While these deep-learning based approaches achieve good performance, they require large training sets and high computing resources. Unlike these studies, we focus on the use of implicit crowdsourcing which requires much less training data and computational cost. In addition to deep learning, there are some domain specific algorithms~\cite{DBLP:conf/mm/RuiGA00,DBLP:journals/tip/EkinTM03}. Unlike these works, we focus on a more general approach. Video summarization~\cite{DBLP:journals/corr/OtaniNRHY16a,DBLP:conf/cvpr/YaoMR16} aims to generate a condensed video to summarize the story of the entire video. Highlight detection often serves as the first step of video summarization and generates a small number of candidate highlights. 

\vspace{.5em}

{\noindent \bf Explicit Crowdsourcing.} There are some works using explicit crowdsourcing for video analysis~\cite{kaspar2018crowd, huang2017leveraging, xu2014users}. That is, they ask crowd workers to do a certain video-related task explicitly, e.g., video segmentation~\cite{kaspar2018crowd}, video tagging~\cite{xu2014users}.  However, none of these studies attempt to apply implicit crowdsourcing to video highlight detection, which is a more monetary-cost efficient and natural way to collect essential data.

\vspace{.5em}

{\noindent \bf Implicit Crowdsourcing (User Comments).} There are some works on the use of user comments~\cite{Wu:2014:CTV:2623330.2623625,ping2017video} for video analysis. A LDA model was proposed to generate video tags from time-stamped comments~\cite{Wu:2014:CTV:2623330.2623625}. Another work uses word embedding to extract highlights from time-stamped comments for movies~\cite{ping2017video}. They are different from \proj in three aspects. (1) They only focus on user commenting data while \proj considers both user commenting data and user viewing behavioral data (see Section~\ref{sec:extract}). (2) They use bag of words or word embedding as features while \proj use more general features (see Section~\ref{subsec:design-choice}). (3) They use time-stamped comments rather than live chat messages, thus they do not face the challenge that there is a delay between video content and the comments (see Section~\ref{subsec:impl}). 

Twitter data has been leveraged to detect events in some studies~\cite{DBLP:conf/mm/XuWWLD06,DBLP:conf/icmcs/HsiehLCH12,DBLP:journals/tkde/SakakiOM13}. However, live chat messages are usually shorter and more noisy thus requiring the development of new techniques.


\vspace{.5em}

{\noindent \bf Implicit Crowdsourcing (User Viewing Behaviors).} HCI and Web researchers have designed systems using click-through or interaction data to measure user engagement. For example, the research on MOOC videos or how-to videos leverage interactions as engagement measurement to detect interesting or difficult parts of videos (e.g.,~\cite{Chorianopoulos2013},~\cite{DBLP:conf/uist/KimGCLGM14}). Some studies have also leveraged interaction data to predict audience's drop-out rate and analyzed the cause of interaction peak~\cite{Kim:2014:UID:2556325.2566237,DBLP:journals/corr/SinhaJLD14}. These works simply sum up all users' watching sessions along the video and get curves between watched frequency and video timestamps. We have compared with these methods, but found that they did not perform well on our collected user interaction data since when users interact with a casual video, their viewing behaviors are much more unpredictable. The experimental results can be found in Section~\ref{subsec:extractor-exp}.

\section{The \proj Workflow}\label{sec:workflow}

\begin{figure*}[t]\vspace{-1.5em}
    \centering
    \includegraphics[scale = 0.5]{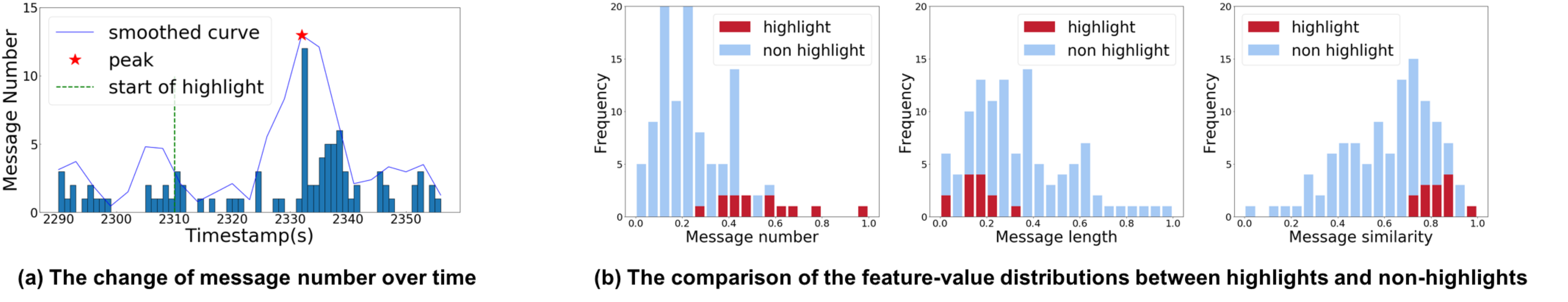}
    \caption{Analysis of the Chat Data in a Twitch Video.}\vspace{-1.5em}
    \label{fig:analyze-chat}
\end{figure*}

Figure~\ref{fig:Overview} depicts the \proj workflow. The workflow consists of two major components: Highlight Initializer and Highlight Extractor. The former determines which part of the video could be a highlight, and the latter identifies the exact boundary (start and end time) of each highlight. We will use a simple example to illustrate how they work as well as the challenges that they face. Consider a one-hour video $\mathcal{V} = [0, 3600]$, which starts at 0s and ends at 3600s. Suppose the video has a highlight  between 1900s and 2005s, denoted by $h = [1990, 2005]$. The goal is to extract $h$ from $\mathcal{V}$.

\vspace{.5em}

{\noindent \bf Highlight Initializer.} Highlight Initializer aims to identify an approximate start position of each highlight so that if a user starts watching the video from this position, she can tell that there is a highlight nearby. For example, 2000 is a good position since it is within the highlight range $[1990, 2005]$ but 2100 is a bad one since it is very far away from the highlight. We observe that most live videos allow users to leave chat messages in real time. This can be leveraged as a kind of implicit feedback. However, time-stamped chat messages are short and noisy, thus it is challenging to use them to implement an accurate Highlight Initializer. We will discuss how to address this challenge in Section~\ref{sec:init}. 

\vspace{.5em}

{\noindent \bf Highlight Extractor.} Suppose the above component returns 2000 as a result. We will add a ``red dot'' at this position on the progress bar of the recorded video (see Figure~\ref{fig:Overview}). A red dot can be seen as a hint, which informs users that there could be a highlight nearby. Users can click on the red dot and start watching the video. They may drag the progress bar backward if they find the video clip interesting and want to watch it again, or they may drag the progress bard forward if they find the video clip uninteresting. All these interactions can be leveraged as another kind of implicit feedback to extract highlights. However, user interactions are noisy. It is challenging to leverage the user interaction data to implement an accurate Highlight Extractor. We will discuss how to address this challenge in Section~\ref{sec:extract}. 

\section{Highlight Initializer} \label{sec:init}
This section presents the design of Highlight Initializer. We first define the design objective, then discuss the design choices, and finally propose the detailed implementation.

\subsection{Design Objective}\label{subsec:obj}
There could be many highlights in a video, but most users are only interested in viewing the top-$k$ ones. Highlight Initializer aims to find an approximate start position for each top-$k$ highlight. 

Next, we formally define what is a \emph{good} approximate start position (i.e., what is a good red dot). The goal is to make users see a highlight shortly after they start watching the video from a red dot. Let $h = [s, e]$ denote a highlight and $r$ denote the red dot w.r.t. $h$. We call $r$ a good red dot if it meets three requirements. 

First, the red dot should not be put after the end of the highlight (i.e., $r \leq e$). Otherwise, a user is very likely to miss the highlight. This is because a user typically clicks the red dot~$r$ and starts watching the video for a short period of time. If nothing interesting happens, she may skip to the next red dot. Second, the red dot should not be put at more than 10s before the start of the highlight (i.e., $r \geq s-10$). Based on existing studies (e.g.,~\cite{response-time}), people can accept less than 10s delay, but may lose their patience when the delay is longer. Third, it is not useful to generate two red dots that are very close to each other. Thus, we require that there does not exist another red dot $r'$ such that $|r-r'| \leq \delta$, where $\delta$ is a system parameter and is set to  120s by default.

With the definition of good red dots, we define the design objective of Highlight Initializer.

\vspace{.5em}

{\noindent \bf Objective.} \emph{Given a recorded live video along with time-stamped messages, and a user-specified threshold $k$, Highlight Initializer aims to identify a set of $k$ \emph{good} red dots.}

\subsection{Design Choices}\label{subsec:design-choice}
We face different choices when designing the Highlight Initializer. We will explain how the decision is made for each choice.

\vspace{.5em}

{\noindent \bf Video vs. Chat Data.} We choose to only use chat data instead of video data to identify red dots. This design choice has two advantages.  First, we can use a small set of training data (e.g., 1 labeled video) to train a good model over chat data. But, it is hard to achieve this for video data. Second, processing video data often requires high computing resources. Since chat data is much smaller in size than video data, this limitation can be avoided. On the other hand, a chat-data based approach may not work well for videos with few chat messages. Nevertheless, as will be shown in the experiment, our model performs well on the videos with 500 chat messages per hour. We find that the majority (more than 80\%) of popular videos in Twitch meet this requirement. For the remaining unpopular videos,  there may not be a strong demand to generate highlights for them. 
\vspace{.5em}

{\noindent \bf General vs. Domain-specific Features.} We seek to build a Machine Learning (ML)  model to identify red dots. There are two kinds of features that can be used by the model. General features are independent of video type (e.g., \dota game vs. \lol game). For example, \emph{message number} can be seen as a kind of general feature because we can extract this feature for any type of video and use it as a strong predictor for highlights.  In contrast, domain-specific features are highly dependent on the selected domains. For example, the keyword ``{\small \textsf{Goal}}'' is a domain-specific feature since it can be used to detect highlights in a Soccer game, but not in a Dota game. We choose to use general features rather than domain-specific features. This will allow our model to have good generalization. 

\subsection{Implementation}\label{subsec:impl}

\sloppy
We implement the Highlight Initializer component based on the above design choices. In the following, we first present a naive implementation and identify its limitations. We then propose our implementation to overcome these limitations. 

\subsubsection{Naive Implementation}
A naive implementation is to count which part of the video has the largest message number and put a red dot at that position.  Figure~\ref{fig:analyze-chat}(a) shows a real-world example. It plots a histogram along with the smoothed curve of the message number in a Twitch live video. We can see that 2332s has the largest message number, thus this naive implementation will put a red dot at 2332s. 

Unfortunately, this implementation does not perform well in practice due to two reasons. The first reason is that having the largest message number does not always mean that users are chatting about a highlight. For instance, there could be advertisement chat-bots which post quite a few messages in a very short period of time. The second reason is that in a live video, users will only chat about a highlight \emph{after} they have seen a highlight.  Thus, there is a delay between the start position of a highlight and its comments. For example, in Figure~\ref{fig:analyze-chat}(a), we can see the delay (the distance between the green dotted line and the red dot) is around 20s.  This naive implementation fails to capture the delay. 
\\

\subsubsection{Our Implementation}
Our implementation consists of two stages.

\vspace{.5em}

{\noindent \bf Prediction.} The prediction stage aims to address the first issue mentioned above. Given a set of chat messages within a short sliding window (e.g., 25s), we build a predictive model to determine whether the messages in the sliding window are talking about a highlight or not. 
We propose three general features for the model.
\begin{itemize}[leftmargin=*]\setlength\itemsep{.15em}
    \item \emph{Message Number} is the number of the messages in the sliding window. The naive implementation only considers this feature.  
    \item \emph{Message Length} calculates the average length of the messages in the sliding window, where the length of a message is defined as the number of words in the message. We observe that if viewers see a highlight, they tend to leave short messages. If their messages are long, they typically chat about something else.
    \item \emph{Message Similarity} measures the similarity of the messages in the sliding window. If the messages are similar to each other, they are more likely to chat about the same topic, instead of random chatting. We use Bag of Words to represent each message as a binary vector and apply one-cluster K-means to find the center of messages. The message similarity is computed as the average similarity of each message to the center. The computation of message similarity can be further enhanced with more sophisticated word representation (e.g., word embedding). 
\end{itemize}

To make these features generalize well, we normalize them to the range in $\left[0,1\right]$ and build a logistic regression model to combine them. We examine the effectiveness of each feature on Twitch chat data. Figure~\ref{fig:analyze-chat}(b) shows the analysis results of a random video. The video contains 1860 chat messages in total. We divide them into 109 non-overlapping sliding windows, where 13 are labeled as highlights and 96 are labeled as non-highlights. For each feature, we compare the feature-value distributions of highlights and non-highlights. We can see that their distributions are quite different. For example, for  the message-length feature, all the highlights are between 0 and 0.4, but non-highlights can be any length. 

\vspace{.5em}

{\noindent \bf Adjustment.} The adjustment stage aims to overcome the second limitation of the naive implementation. Given a set of messages in a sliding window which are predicted to be talking about a highlight, we aim to estimate the start position of the highlight.

The key observation is that people can only comment on a highlight after they have seen it. We first detect the \emph{peak} in the sliding window, where a peak represents the time when the message number reaches the top. After that, we train a model to capture the relationship between the peak's position ($\timepeak$) and the highlight's start position ($\timestart$). 

The current implementation considers a simple linear relationship, i.e., $\timestart = \timepeak - c$, where $c$ is a constant value. We can learn the optimal value of $c$ from training data. Specifically, for each labeled highlight $i$, the highlight's ground-truth start position is denoted by $\timestart_i$. Since it is predicated as $\timepeak_i - c$, the red dot will be put at $\timepeak_i - c$.  Our goal is to identify as many good red dots as possible. Thus, we aim to find the best $c$ such that
\begin{equation*}\underset{c}{\operatorname{arg\,max}} \sum_i \emph{\textrm{reward}}(\timepeak_i - c,~ \timestart_i),
\end{equation*}
where $\emph{\textrm{reward}}(\cdot) = 1$ if it is a good red dot; $\emph{\textrm{reward}}(\cdot) = 0$, otherwise.


Once $c$ is obtained, we can use it to get the red dot positions. For example, suppose the learned $c$ = 20s. It means that we will move the peak backward by 20s. Imagine $\timepeak$ = 2010s. Then we will select $\timepeak$ - 20s = 1990s as a red dot's position. This simple linear relationship leads to good performance as shown in later experiments. We defer the exploration of more complex relationships to future work.

\vspace{.5em}

\vspace{.5em}
{\noindent \bf Algorithm Description.} To put everything together, Algorithm~\ref{machineStage} shows the pseudo-code of the Highlight Initializer component. The input consists of the set $M$ of all the time-stamped messages of video $v$, the video length $t$, the number of one's desired highlights~$k$, the sliding window size~$l$, the adjustment value~$c$, and Trained Logistic Regression Model, $\emph{\textrm{LRmodel}}$. The output is the highlight sliding window list, $H = \lbrace{(s_j,e_j)~|~ j = 0,..., k-1}\rbrace$, where $(s_j, e_j)$ is respectively the start and end time of a sliding window $j$.

In line 1, we initially generate the sliding window list $W = \lbrace{(s_i,e_i)~|~ i = 0,..., n}\rbrace$. When two sliding windows have an overlap, we keep the one with more messages. From line 2 to line 6, for each sliding window $w_i = (s_i, e_i)$, we apply the trained logistic regression model on the feature vector $f_i = (num_i, len_i, sim_i)$ which is extracted from the massages whose timestamps are in the range of $(s_i, e_i)$. Lines 7 and 8 retrieve the top-k highlight sliding windows $H$. In $Top$ function, we make sure that $H$ does not contain too close highlights. From line 9 to 11, we adjust the start time by $c$ for each sliding window in $H$. Finally, we return $H$ as an output.

\setlength{\textfloatsep}{1em}
\begin{algorithm}[t] \small
  \caption{Highlight Initializer for one video $v$.}
  \label{machineStage}
  \DontPrintSemicolon
  \SetAlgoLined
  \SetKwInOut{Input}{Input}\SetKwInOut{Output}{Output}
  \Input{$M$: all the messages; $t$: video length; $k$: \# of desired highlights; $l$: sliding window size; $c$: adjustment value;  $\emph{\textrm{LRmodel}}$: trained logistic regression model}
  \Output{$H$: top-$k$ sliding window list.}
  \BlankLine
  $W \gets$ get\_sliding\_wins($M,l$)\;
  \ForEach{\emph{sliding window} $W[i]$}{
      $f_i \gets$ Feature\_vec($W[i],M$) \tcp{\small \textrm{\footnotesize \emph{Normalized f=(num,~len,~sim)}}}
      $p_i \gets$ $\emph{\textrm{LRmodel}}$.predict$(f_i)$ \tcp{\small \textrm{\footnotesize\emph{Get predicted probability}}}
      $W[i] \gets W[i].\textrm{append}(p_i)$\;
  }
  $W_{\textrm{sorted}} \gets $Sort $W$ by $p$ \;
  $H\gets$Top($k,W_{\textrm{sorted}}'$)\;
  \ForEach{\emph{highlight window} $H[j]$}{ 
      $H[j]\gets(s_j - c, e_j)$ \tcp{\footnotesize \textrm{Adjustment}}
  }
  \Return{$H$}
\end{algorithm}

\section{Highlight Extractor}\label{sec:extract}

This section presents the design of Highlight Extractor. We first define the design objective, then discuss the challenges, and finally propose the detailed implementation.     

\subsection{Design Objective}
Highlight Extractor aims to identify the boundary (start and end positions) of each highlight using user interaction data.

\vspace{.25em}

{\noindent \bf  User Interaction Data.} While watching a video, a user may have different kinds of interactions with the video (e.g., Play, Pause, Seek Forward, and Seek Backward). We analyze user interaction data, and find that if a certain part of video has a large number of views, then this part is very likely to be a highlight. Based on this observation, we transform user interaction data into \emph{play data}, where each record is in the form of: $\langle user , \watch(s, e)\rangle$. For example, $\langle \textrm{Alice} , \watch(100, 120)\rangle$ means that the user Alice starts playing the video at 100s, and stops at 120s. If the context is clear, we will abbreviate it as $\watch(s, e)$ and call $\watch(s, e)$ a \emph{play}.

We leverage the play data to extract highlights. Note that if a play is far away from a red dot, it may be associated with another highlight. Thus, we only consider the plays within $[-\Delta, \Delta]$ around a red dot ($\Delta = 60$s by default). 

The following defines the objective of Highlight Extractor.

\vspace{.5em}

\sloppy
{\noindent \bf Objective.} \emph{Given the play data  $\watch(s_1, e_1)$ $,\cdots$ $,\watch(s_n, e_n)$ w.r.t. a red dot, Highlight Extractor aims to identify the start and end positions of the highlight of the red dot.}

\subsection{Challenges}
To achieve the objective, we need to address the following challenges.
\vspace{.5em}

{\noindent \bf How to filter play data?} Play data could be very noisy. For example, a user may randomly pick up a position $s$, and watch for a few seconds (e.g., 5s) to check whether this part of video is interesting or not. If uninteresting, she may jump to another position. Obviously, we should filter this $\watch(s, s+5)$ since it cannot be interpreted as the user enjoying watching $[s, s+5]$. Note that this example only shows one type of noisy play. There could be many others that need to be considered. 

\vspace{.5em}

\sloppy

{\noindent \bf How to aggregate play data?} Let $\watch(s'_1, e'_1),$ $\cdots,$ $\watch(s'_m, e'_m)$ donate the play data after the filtering. Each play can be considered as a vote for the highlight. For example, $\watch(1990, 2010)$ means that the user votes 1990s and 2010s as the start and end positions of the highlight. Users may have different opinions about the highlight. We can aggregate their opinions using {\small \textsf{median}} because it is robust to outliers. Thus, the new start and end positions are computed as ${\small \textsf{median}}(s'_1, s'_2, \cdots, s'_m)$ and ${\small \textsf{median}}(e'_1, e'_2, \cdots, e'_m)$. 

\fussy

Unfortunately, when applying this idea to real-world user interaction data, it does not always work well. We have a very interesting observation: whether this idea works well or not strongly depends  on the relative positions of the red dot and the highlight. There are two possible relative positions:

\textit{Type \Rmnum{1}}: the red dot is \emph{after} the end of the highlight; 

\textit{Type \Rmnum{2}}: the red dot is \emph{before} the end of the highlight. 

Since many users start watching the video from a red dot, if they do not find anything interesting, they may skip to the next red dot. Imagine the red dot is put after the end of the highlight (i.e., \textit{Type~\Rmnum{1}}). Many users may miss the highlight, thus their play data are not reliable indicators of the highlight. Imagine the red dot is put before the start of the highlight (i.e., \textit{Type~\Rmnum{2}}). Many users will watch the same highlight, thus their play data follow a similar pattern. 

\begin{figure}[t]
    \begin{subfigure}[b]{0.225\textwidth}
        \includegraphics[width=\linewidth]{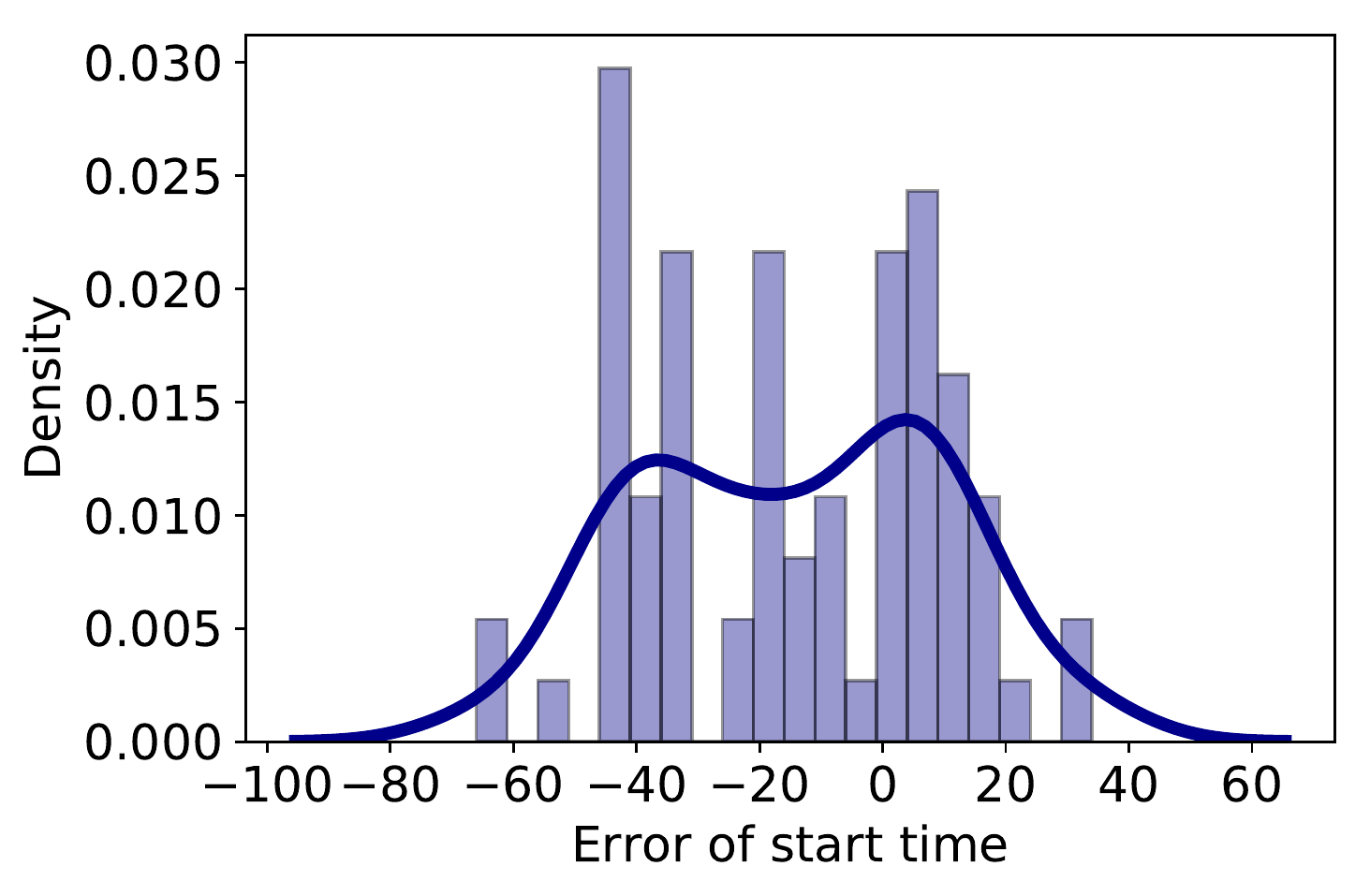}\vspace{-.5em}
        \subcaption{Type \Rmnum{1}}
        \label{fig:type1_error}
    \end{subfigure}
    \begin{subfigure}[b]{0.225\textwidth}
        \includegraphics[width=\linewidth]{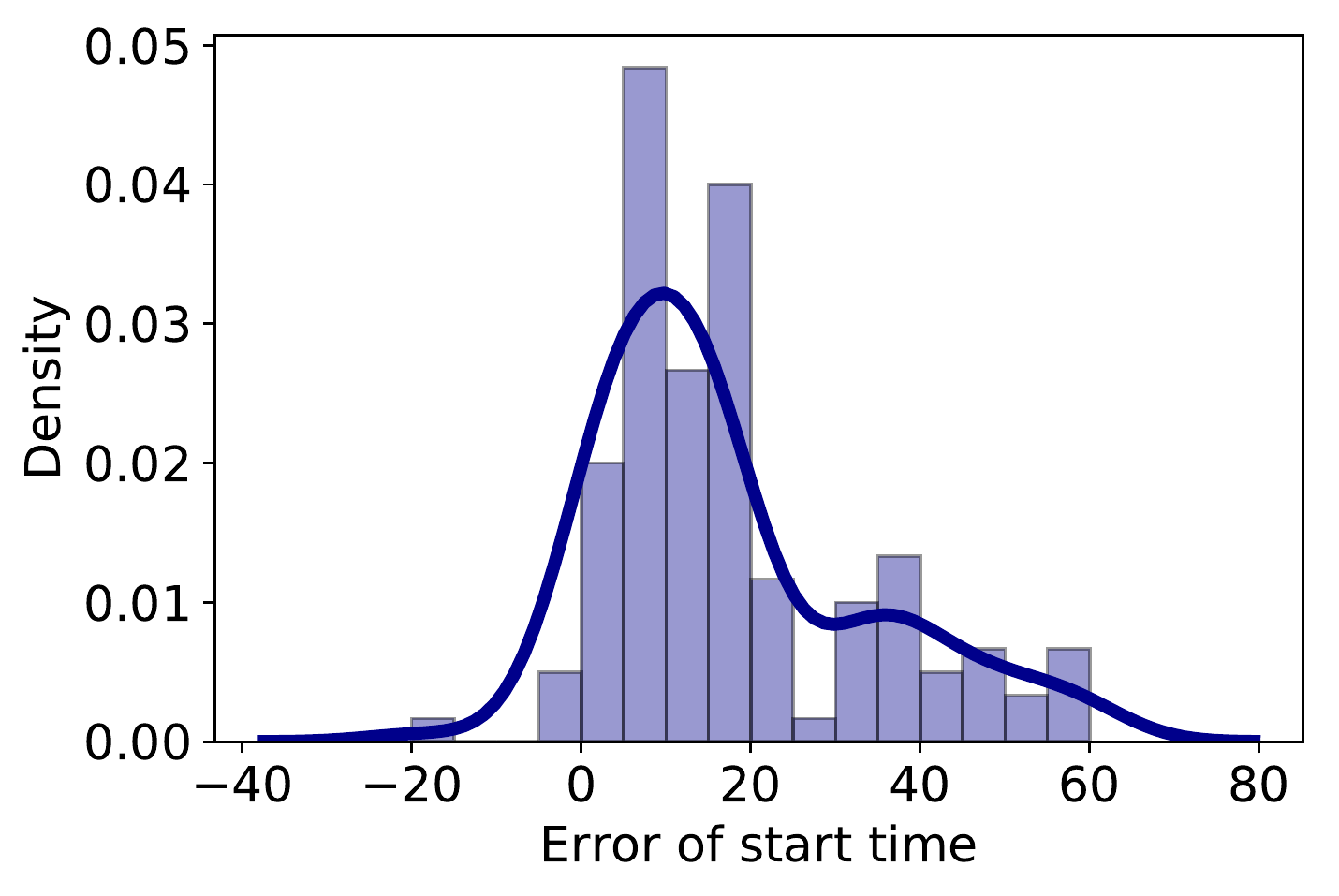}\vspace{-.5em}
        \caption{Type \Rmnum{2}}
        \label{fig:type2_error}
    \end{subfigure}
    \caption{Distribution of the difference between each play's start position and the ground-truth start position.}
    \label{fig:type}\vspace{-.5em}
\end{figure}

To further examine this observation, we calculate the difference of each play's start position and the ground-truth start position.  Figure~\ref{fig:type}(a) shows the distribution of all plays of \textit{Type~\Rmnum{1}}. We can see the curve approximately follows a uniform distribution between -40 and +20. It shows that the play activities for \textit{Type~\Rmnum{1}} are quite diverse. Users may either play back randomly in order to find the highlight or skip to the next highlight. In comparison, Figure~\ref{fig:type}(b) shows the distribution of all plays of \textit{Type~\Rmnum{2}}. We can see the curve approximately follows a normal distribution. It implies that most plays for \textit{Type~\Rmnum{2}} correspond to highlight watching. 

This observation poses two new questions. The first one is that given a red dot, how to determine whether it belongs to \textit{Type~\Rmnum{1}} or \textit{Type~\Rmnum{2}}? The second one is that after a red dot is classified as \textit{Type~\Rmnum{1}} or \textit{\Rmnum{2}}, how to aggregate its play data?



\vspace{-.5em}
\subsection{Implementation}

We propose a series of techniques to address these challenges. The following shows our detailed implementation.

\vspace{.5em}

{\noindent \bf Filtering.} The main idea is to filter the plays that are not about watching the highlight but about doing something else (e.g., looking for a highlight).  We observe that if a play is far away from the red dot, it typically does not cover the highlight. Thus, we remove such plays from the data. We also notice that if a play is too long or too short, it tends to have little value. A too short play could indicate that viewers watch for a few seconds and find it uninteresting, while a too long play means that viewers may be watching the entire video. Thus, we remove such plays from the data. Third, there could be some outliers, i.e., the play that is far away from the other plays. We adopt the following outlier detection method to find the outliers and then remove them. 

We construct an undirected graph $G=(V, E)$. Each node $v_{i}=(s_{i}, e_{i}) \in V$ represents a play record, where $(s_i, e_i)$ is respectively the start and end time. Each edge $e_{v_{i},v_{j}} \in E$ represents that $v_{i}$ and $v_{j}$ have overlapping part. Then we find the center node $o \in V$ of the graph which has the largest degree. We select $o$ and its neighbor nodes as valuable play data and treat the others as outliers. That is, \textsf{\small Outliers} = $\{v \in V \mid v \neq o \textrm{ and } e_{v, o} \notin E\}$.

\sloppy
\vspace{.5em}

{\noindent \bf Classification.} Given a red dot, we build a classification model to determine whether it belongs to  \textit{Type \Rmnum{1}} or \textit{Type \Rmnum{2}}. We need to classify the relative position between the red dot and the end of the highlight into \textit{Type \Rmnum{1}} or \textit{Type \Rmnum{2}}. We find that this (unknown) relative position has a strong correlation with the (known) relative position between the red dot and observed play data. Therefore, we identify the following three features.

\begin{itemize}[leftmargin=*]\setlength\itemsep{-.15em}
    \item \emph{\#~Plays after red dot} computes the number of plays which start at or after the red dot.  
    \item \emph{\#~Plays before red dot} computes the number of plays which end before the red dot. 
    \item \emph{\#~Plays across red dot} computes the number of plays which starts before the red dot and ends after the red dot. 
\end{itemize}

\begin{figure}[t]\vspace{-2em}
    \centering
    \includegraphics[scale = 0.45]{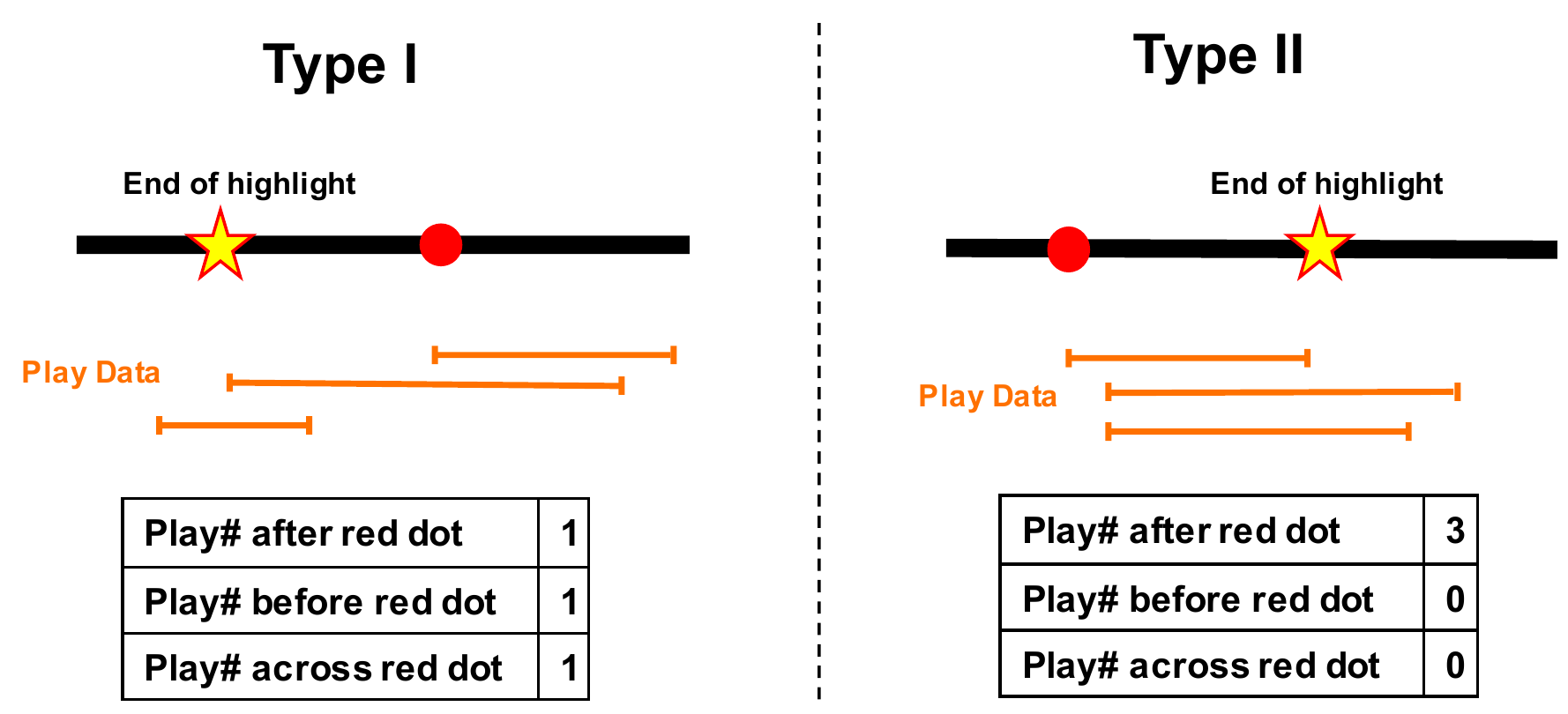}
    \caption{An illustration of three features for classifying the relative position between red dot and the end of highlight.}
    \label{fig:classification}
\end{figure}

Figure~\ref{fig:classification} shows an example to illustrate the three features. For \textit{Type \Rmnum{1}}, since the highlight ends before the red dot,  some users play before or across the red dot in order to find the highlight. In comparison, there is no such play in \textit{Type \Rmnum{2}} since if a user starts watching the video at the red dot, she will see the highlight.  Our experiments show that our classification model achieves high accuracy (around 80\%).

\begin{algorithm}[t]\small
    \caption{Highlight Extractor for one highlight $h$}
    \label{CrowdStage}
    \DontPrintSemicolon
    \SetAlgoLined
    \SetKwInOut{Input}{Input}\SetKwInOut{Output}{Output}
    \Input{Highlight $h = (s ,e)$;\\ Moving duration for \textit{Type \Rmnum{1}}, $m$.}
    \Output{Highlight refined boundaries, $h = (s', e')$.}
    \BlankLine
    \Repeat{\emph {Red dot position converge}}{
        $I\gets$ get\_interact()\;
         
        $P\gets$ filter($I$) \tcp{\textrm{\small \emph{Filtering}}}
        
        $f\gets$ feature\_vec($P$) \tcp{\small \textrm{\emph{f=(before\_red, after\_red, across\_red)}}}
        $label \gets$ classification($f$) \tcp{\textrm{\small \emph{Classification}}}
        \tcp{\textrm{\small \emph{Aggregation}}}
        \eIf{$label = \textrm{Type}~\Rmnum{2}$}{
            \ForEach{$p$ in $P$}{
                \uIf{p.e < h.s}{Remove(p) \tcp{\small \textrm{\emph{Remove plays before red dots}}}}
            }
            $h.s' \gets$ median($P.s$) ; $h.e' \gets$ median($P.e$) \;
        }{
            $h.s' \gets h.s - m$\;
        }
    }
    \Return{$h$}
\end{algorithm}

\vspace{.5em}

{\noindent \bf Aggregation.} Different aggregation strategies are proposed for \textit{Type \Rmnum{1}} and \textit{Type \Rmnum{2}}, respectively. 

For \textit{Type \Rmnum{2}}, as Figure~\ref{fig:type}(b) shows, the play patterns of most users are similar. The median of the start time offsets is between 5 and 10. This is because that the most exciting part of the highlight usually happens a few seconds after its start point, which causes users to skip the beginning of the highlight. This kind of error is tolerable. Therefore, we can use \textsf{median} to aggregate their play data.

For \textit{Type \Rmnum{1}}, as seen from Figure~\ref{fig:type}(a), the distribution of start time offsets is rather random. Therefore, we need to collect more play data. Our main idea is to convert a red point from \textit{Type~\Rmnum{1}} to \textit{Type~\Rmnum{2}}. Given that \textit{Type~\Rmnum{2}} can collect high-quality play data, once a red dot is converted to \textit{Type~\Rmnum{2}}, we can get high-quality play data as well. Specifically, once a red dot is classified as \textit{Type~\Rmnum{1}}, we will move it backward by a constant time (e.g., 20s) and collect new interaction data based on the new red dot location. If the new red dot is classified as \textit{Type~\Rmnum{2}}, we apply the \textit{Type~\Rmnum{2}}'s aggregation approach; otherwise, we move it backward by another 20s.

\vspace{.5em}

{\noindent \bf Algorithm Description.} To put everything together, Algorithm \ref{CrowdStage} shows the pseudo-code of the entire Highlight Extractor component. The input consists of a highlight $h = (s, e)$, and a moving duration $m$ for \textit{Type~\Rmnum{1}}, which is a constant described above to convert a \textit{Type~\Rmnum{1}} to a \textit{Type~\Rmnum{2}}. The output is the updated $h=(s',e')$. 

In lines 2 and 3, we get the user interactions $I$ for the current $h$ and filter them to get a list of plays, $P$. In lines 4 and 5, we extract the feature $f$ from $P$ and perform the binary classification to decide $h$'s $label$. From lines 6 to 14, as we describe above in \textsf{\small Aggregation}, we update $h=(s',e')$. If $label$ is \textit{Type {\Rmnum{2}}}, it means the red dot is before the end of the highlight. From lines 7 to 10, we remove the plays whose ends are before the red dot. Then, we calculate the median to update $h$. If $label$ is \textit{Type {\Rmnum{1}}}, it means the red dot is after the end of the highlight, and we move $h.s$ backward by $m$. We iterate this procedure until the red dot position is stable (e.g., $|h.s-h.s'| < \epsilon$).

\section{Deploy \proj in Practice}\label{sec:deploy}

In this section, we discuss two ways to deploy the \proj workflow: one is to wrap it as a web browser extension and the other is to integrate it into existing live streaming platforms.  

\subsection{Web Browser Extension}

\begin{figure}[t]\vspace{-2em}
    \centering
    \includegraphics[scale = 0.3]{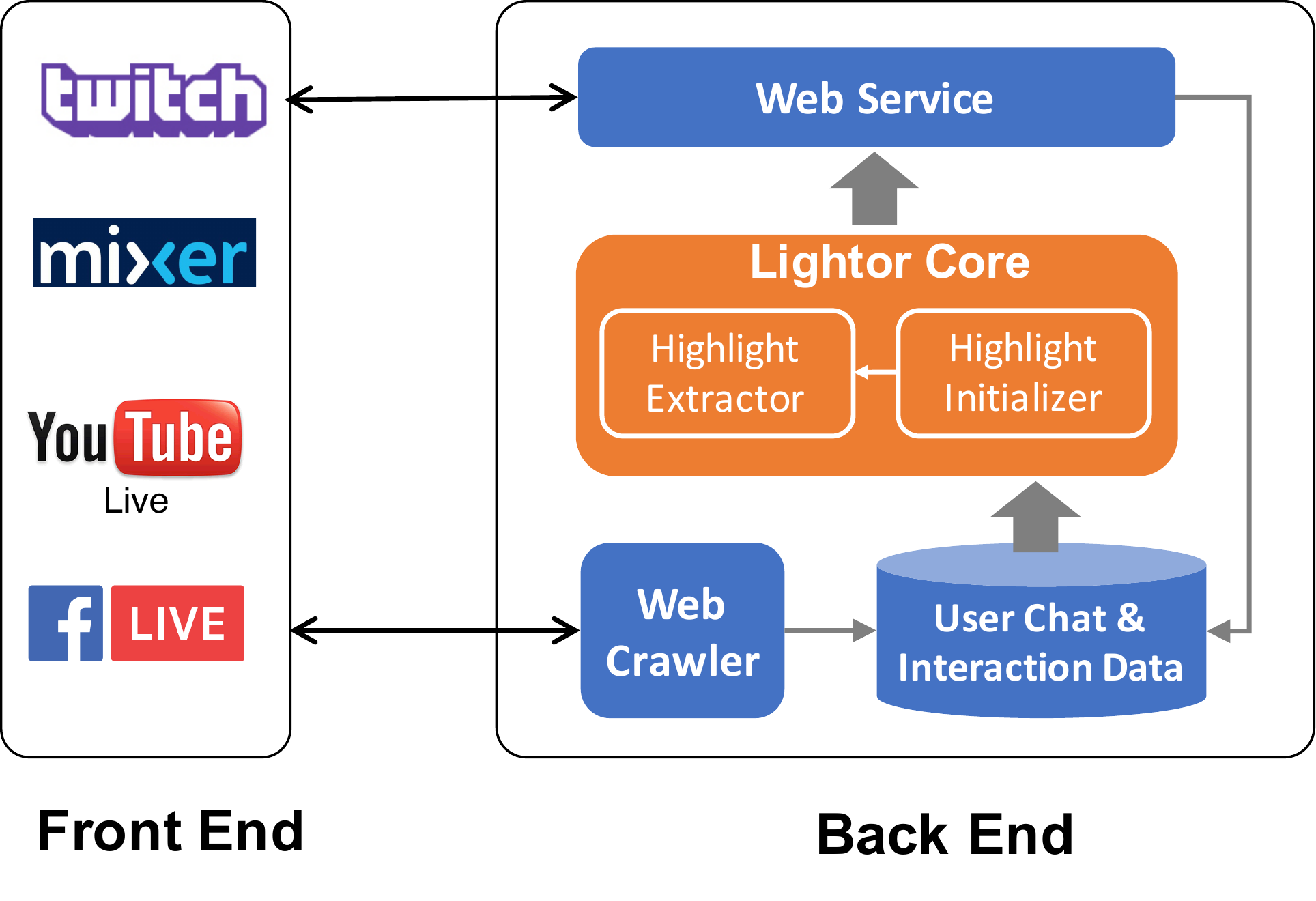} 
    \caption{\proj Web Browser Extension.}
    \label{fig:Extension}
\end{figure}

Figure~\ref{fig:Extension} depicts the architecture of our web browser extension. It has the potential to support any live streaming platform. We will use Twitch as an example to illustrate how it works.

In addition to the  \proj's core components, we need two additional components: Web Service and Web Crawler.

\vspace{.25em}

{\noindent \bf Web Service.} When a user opens up a web page in Twitch, if the URL of the web page is a recorded video, \proj will be automatically activated. It extracts the Video ID from the web page and sends it to the back end server. The server receives the Video ID and checks whether the video's chat messages have been crawled and stored in the database. If not, it will call the web crawler component to crawl the chat messages. After that, it will use the chat data to identify the positions of top-$k$ highlights and return them to the front end. The returned results will be rendered on the original web page by adding red dots on the progress bar of the video. Meanwhile, the user interaction data will be logged. Highlight Extractor will  use the data to refine the results. The refined results will be stored in the database continuously.

\vspace{.25em}

{\noindent \bf Web Crawler.} The web crawler component crawls the chat messages of recorded videos in Twitch. The crawled chat messages will be stored into the database. The crawling process can be executed both \emph{offline} and \emph{online}. The offline crawling periodically checks  a given list of popular channels. If new videos are uploaded in those channels, their chat messages will be crawled accordingly. The online crawling will crawl the chat messages on the fly. It will be triggered if the chat messages of a video do not exist in the database.

\subsection{Integrate Into Live Streaming Platforms}

\proj can be easily deployed into existing live streaming platforms. The only change is to add red dots to the progress bar of recorded videos. It is easy to implement this feature from a technical point of view. Moreover, based on our user study, this new interface is more attractive since it can help users find more interesting highlights.

\proj is also useful for improving existing features. For example, Twitch allows broadcasters to cut and upload the highlights of their recorded videos manually. \proj can provide broadcasters with a set of  highlight candidates. This will help broadcasters save a lot of time when they need to edit their highlights repeatedly.

\section{Experiments} \label{sec:exp}
We evaluate \proj on real live video data. The experiments aim to answer three questions. (1) How well does Highlight Initializer perform? (2) How well does Highlight Extractor perform? (3) How does \proj compare with deep-learning based approaches? We will first describe experimental settings and then present  experimental results.
We provide a reproducibility report in the github repo (\url{https://github.com/sfu-db/Lightor_Exp}), including  data description,  data preprocessing, model parameters and configurations,  and a Jupyter notebook to reproduce all experimental figures.

\subsection{Experimental Settings}
Game videos dominate mainstream live streaming platforms such as Twitch and Mixer. They were also used to evaluate the state-of-the-art highlight detection approaches~\cite{DBLP:conf/emnlp/FuLBB17,DBLP:journals/corr/Song16a}. Therefore, we evaluated \proj using two popular games from Twitch: \dota and \lol. 

\vspace{.25em}

{\noindent \textbf{Game Videos.}} (1) \dota. We crawled 60 live videos in Dota~2 using Twitch APIs. The length of each video is between 0.5 hour to 2 hours. We asked experienced game players to watch each video and manually label the start and end positions of each highlight. Each video contains 10 labeled highlights on average. The length of each highlight is between 5s to 50s.   (2) \lol. We  selected 173 live videos of League of Legends (LoL) from NACLS dataset~\cite{DBLP:conf/emnlp/FuLBB17}. The length of each video is between 0.5 hour to 1 hour. The labels were obtained by matching with highlight collections of a YouTube channel. Each video contains 14 labeled highlights on average. The length of each highlight is between 2s to 81s.

The \dota and \lol datasets are different in two aspects. First, the game types are different, and thus raw visual and textual features do not generalize well. Second, the \dota videos were from Twitch personal channels, but the \lol videos came from North America League of Legends Championship Series. Thus, their chat data have different characteristics.  

\vspace{.25em}

{\noindent \textbf{Chat and Play Data.}} \proj relies on two kinds of user data: chat data and play data. For chat data, Live streaming platforms make the data accessible. We used their APIs to crawl the data. The number of chat messages crawled for each video is between 800 to 4300. 

It is quite challenging to collect play data since they are not accessible from a live streaming platform. To the best of our knowledge, there is even no  play data publicly available. To collect the data, we recruited game fans from Amazon Mechanical Turk (AMT), and asked them to watch the recorded live videos. Each video's progress bar has a single red dot since we would like to get rid of the influence of nearby red dots and study user interactions on one red dot directly. We collected the user interaction data and then generated the play data. Note that we did not ask the crowd to enter the start and end positions of a highlight. Therefore, the crowd provided us with the boundary of a highlight \emph{implicitly}.  There were 492 workers participating our experiments and we spent about \$750 to create the dataset. We have published the dataset in the above github repo.

\vspace{.5em}

{\noindent \bf Evaluation Metrics:} We used \precisionK to evaluate the performance since most users are only interested in watching a small number of highlights (e.g., k = 5 to 10).  We defined three \precisionK metrics in the experiments.

(1) \chatprecisionK is to evaluate the effectiveness of the prediction stage in Highlight Initializer. The prediction stage sorts chat-message sliding windows based on how likely they are talking about a highlight, and returns the top-$k$ sliding windows. \chatprecisionK  is defined as the percentage of correctly identified sliding windows out of the k identified sliding windows. 

(2) \videoprecisionstartK is to evaluate the precision of the identified start positions of highlights. Since people typically cannot tolerate more than 10s delay, we say a start position $x$ is correct if there exists a highlight $h = [s, e]$ such that $x \in [s-10, e]$.   \videoprecisionstartK is defined as the percentage of correctly identified start positions out of the k identified start positions.

(3) \videoprecisionendK is to evaluate the precision of the identified end positions of highlights. It is  similar to \videoprecisionstartK. We say an end position $y$ is correct if there exists a highlight $h = [s, e]$ such that $y \in [s, e+10]$.  \videoprecisionendK is defined as the percentage of correctly identified end positions out of the k identified end positions.

\vspace{.5em}

{\noindent \textbf{Baselines:}} Highlight detection is a multidisciplinary problem and there are many solutions proposed in different areas. We compared \proj with the following approaches. 

\begin{itemize}[leftmargin=*]\setlength\itemsep{0.25em}
    \item \emph{Social network based methods} analyze online social network data to detect events. \toretter\cite{DBLP:journals/tkde/SakakiOM13} is an earthquake detection system based on Twitter Data. We applied it to our chat data to detect the start of each highlight.  We compared it with \proj's Highlight Initializer in Section~\ref{subsec:initializer-exp}.
    \item \emph{Online learning based methods} aim to detect key clips of How-To and lecture videos by analyzing viewers' viewing behaviors. \socialskip\cite{Chorianopoulos2013} analyzes viewers' skipping and jumping backwards interactions while \moocer\cite{Kim:2014:UID:2556325.2566237} mainly focuses on normal playing interactions. We applied them to our play data to identify the boundary of each highlight. We compared these two baselines with \proj's Highlight Extractor in Section~\ref{subsec:extractor-exp}.
    \item \emph{Deep Learning based methods} train a deep neural network model on a large amount of labeled data and then apply the model to classify each video frame as highlight or non-highlight.   We compared \proj with the state-of-art deep learning approach~\cite{DBLP:conf/emnlp/FuLBB17} in Section~\ref{subsec:deeplearning}. The approach has two models, where chat model (\chatlstm) is trained on chat data only and chat-video joint model (\jointlstm) is trained on both chat and video data.
\end{itemize}

\vspace{.5em}

\noindent \textbf{Software Versions and Hardware Configuration:} \proj was implemented using Python 3.5. Logistic regression models were trained using scikit-learn 0.20. The experiments were run over a Ubuntu virtual server with an Intel Xeon CPU E7-4830 v4@2.00GHz with RAM 53GB. The sliding window size was set to 25s. The deep learning models were implemented in PyTorch 1.1 and trained on 4 Nvidia Tesla v100 GPUs.

\vspace{.5em}

In the following, we first evaluate the Highlight Initializer and Highlight Extractor of \proj, and then examine the applicability of \proj on the Twitch platform. Finally, we compare \proj with the deep learning methods.

\subsection{Evaluation of Highlight Initializer}\label{subsec:initializer-exp}
Highlight Initializer consists of prediction and adjustment stages. We evaluated their performance on \dota data.

\begin{figure}[t]\vspace{-2.5em}
        \begin{subfigure}[b]{0.235\textwidth}
                \includegraphics[width=\linewidth]{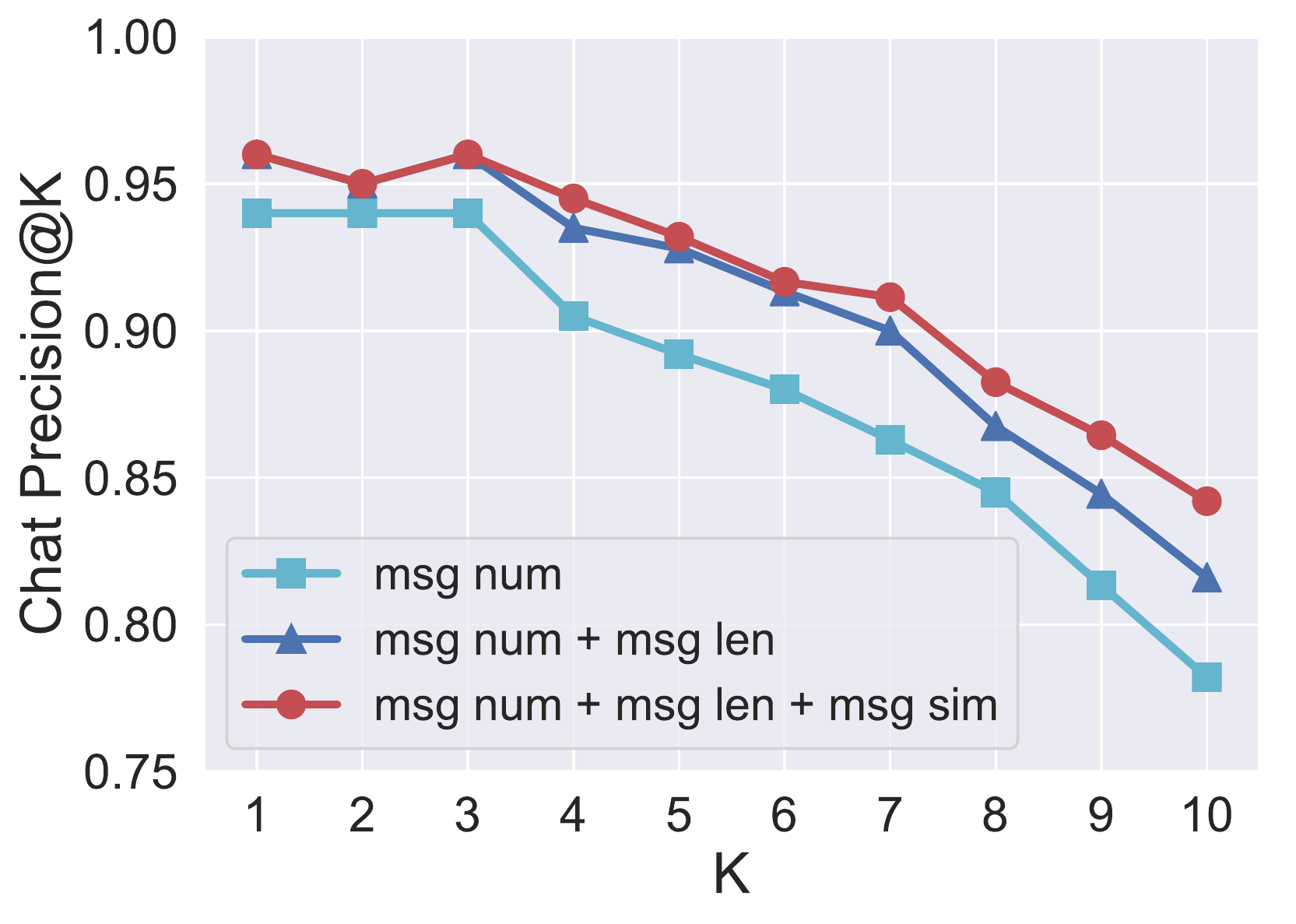}\vspace{-0.5em}
                \caption{Prediction Performance}
                \label{fig:predict}
        \end{subfigure}
        \begin{subfigure}[b]{0.235\textwidth}
                \includegraphics[width=\linewidth]{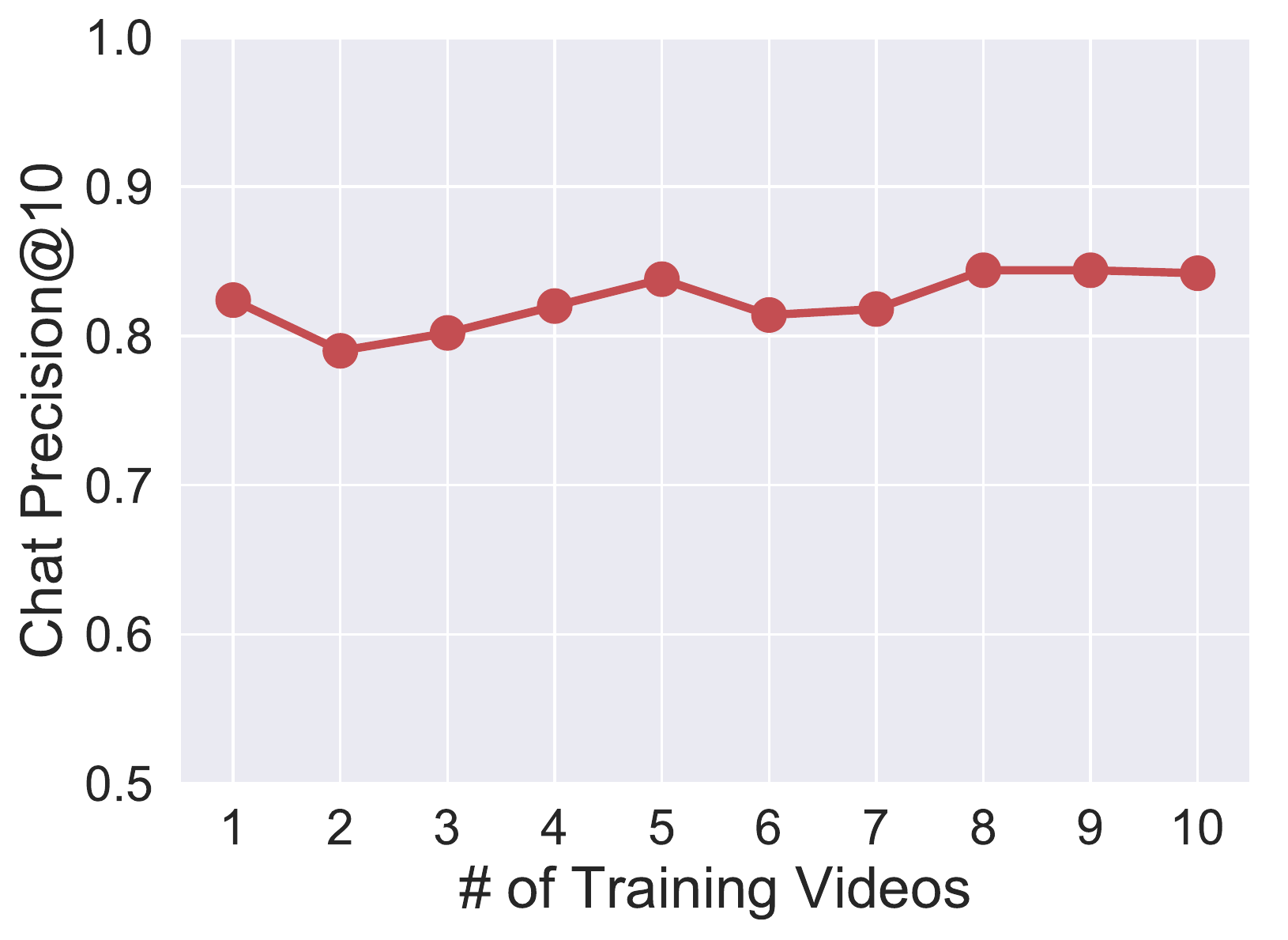}\vspace{-0.5em}
                \caption{Effect of Training Size}
                \label{fig:trainsize_lightor}
        \end{subfigure}
        \caption{Evaluation of Prediction Stage.}
        \label{fig:initializer-predict}\vspace{-.5em}
\end{figure}

\vspace{.25em}

{\noindent \bf Prediction Stage.} The prediction stage is designed to get top sliding windows corresponding to highlights. We propose three features, message number (\emph{msg num}), message length (\emph{msg len}), and message similarity (\emph{msg sim}), and build a logistic regression model based on them. To evaluate the effectiveness of the proposed features, we build two additional logistic regression models using \emph{msg num} and \emph{msg num + msg len}. We used 10 video's sliding windows as training data and used 50 videos' sliding windows as test data. 

Figure~\ref{fig:predict} shows the average \chatprecisionK of the 50 testing videos on different number of desired highlights. We have two interesting observations. First, \emph{msg num} was an effective feature for small k ($\leq 3$) but did not perform well as k got larger (e.g., $k = 10$). This is because that as k increased, it would be more and more challenging to detect new highlights. If we only used  the \emph{msg num} feature, these messages sometimes were sent because viewers were discussing something on random topics which were not related to the highlights. Second, the ML model using all three features was better at capturing the nature of highlight messages especially when one wants to detect more than 5 highlights. The reason is that when viewers saw a highlight, their messages tended to be in a similar pattern. In addition to actively sending more messages, they would send more short messages such as Emojis or Stickers which make the average length of messages in the sliding windows shorter than common ones. When viewers were talking about something particular in the highlights, the messages would have a higher similarity.

We examine whether our method can still perform well on small training data. We varied the number of training videos from 1 to 10 and trained ten models w.r.t. different training size.  Figure~\ref{fig:trainsize_lightor} shows the average \chatprecisionten of each model evaluated on the 50 testing videos. We can see that the performance stayed stable. For example, when there was only one training video, our method can still achieve the precision of 0.82. The reason for this impressive result is that our ML model was built on a small number of highly effective features. 

\begin{figure}[t]\vspace{-2.5em}
        \begin{subfigure}[b]{0.235\textwidth}
                \includegraphics[width=\linewidth]{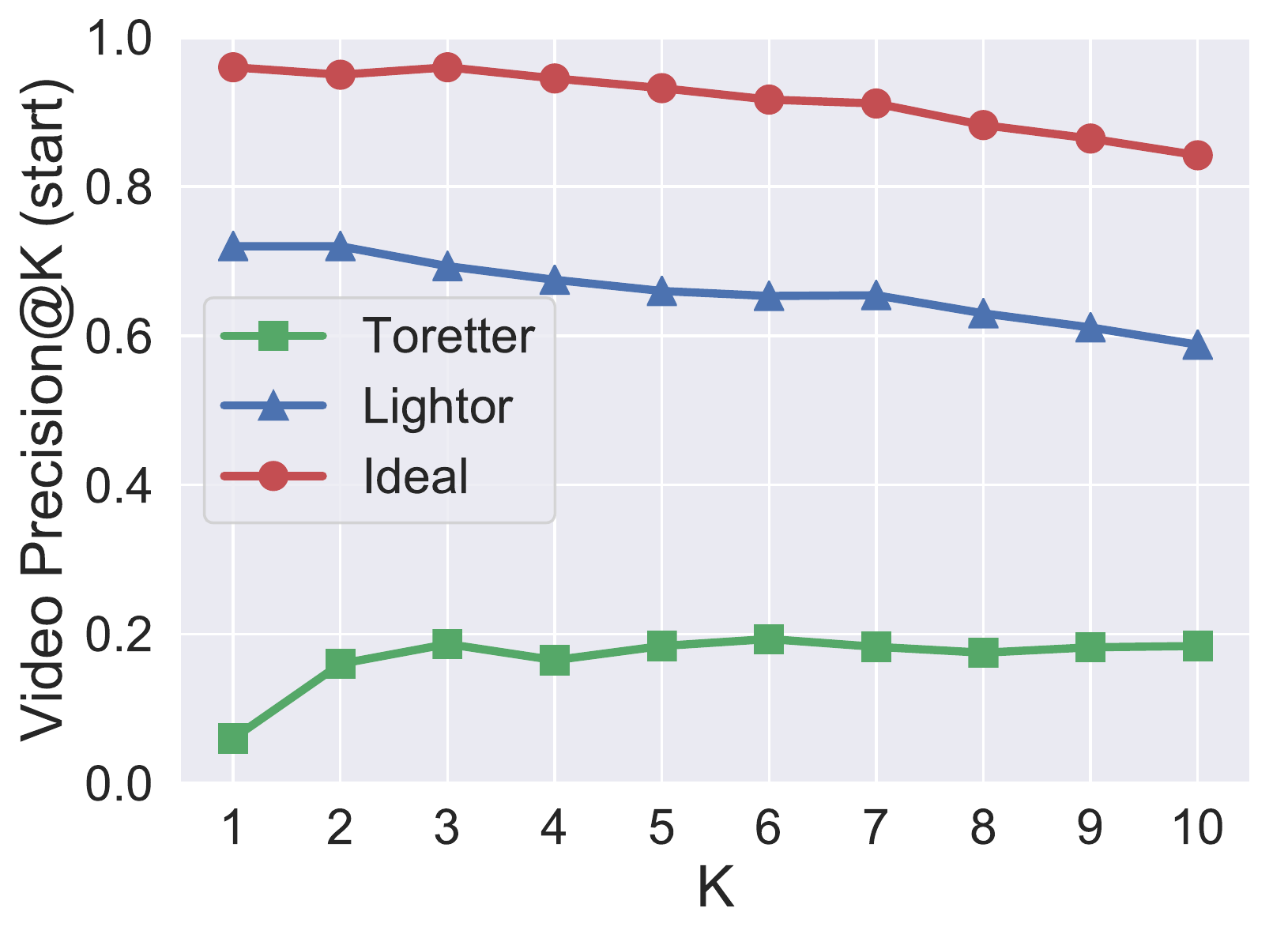}\vspace{-0.5em}
                \caption{Adjustment Performance}
                \label{fig:adjust}
        \end{subfigure}
        \begin{subfigure}[b]{0.235\textwidth}
                \includegraphics[width=\linewidth]{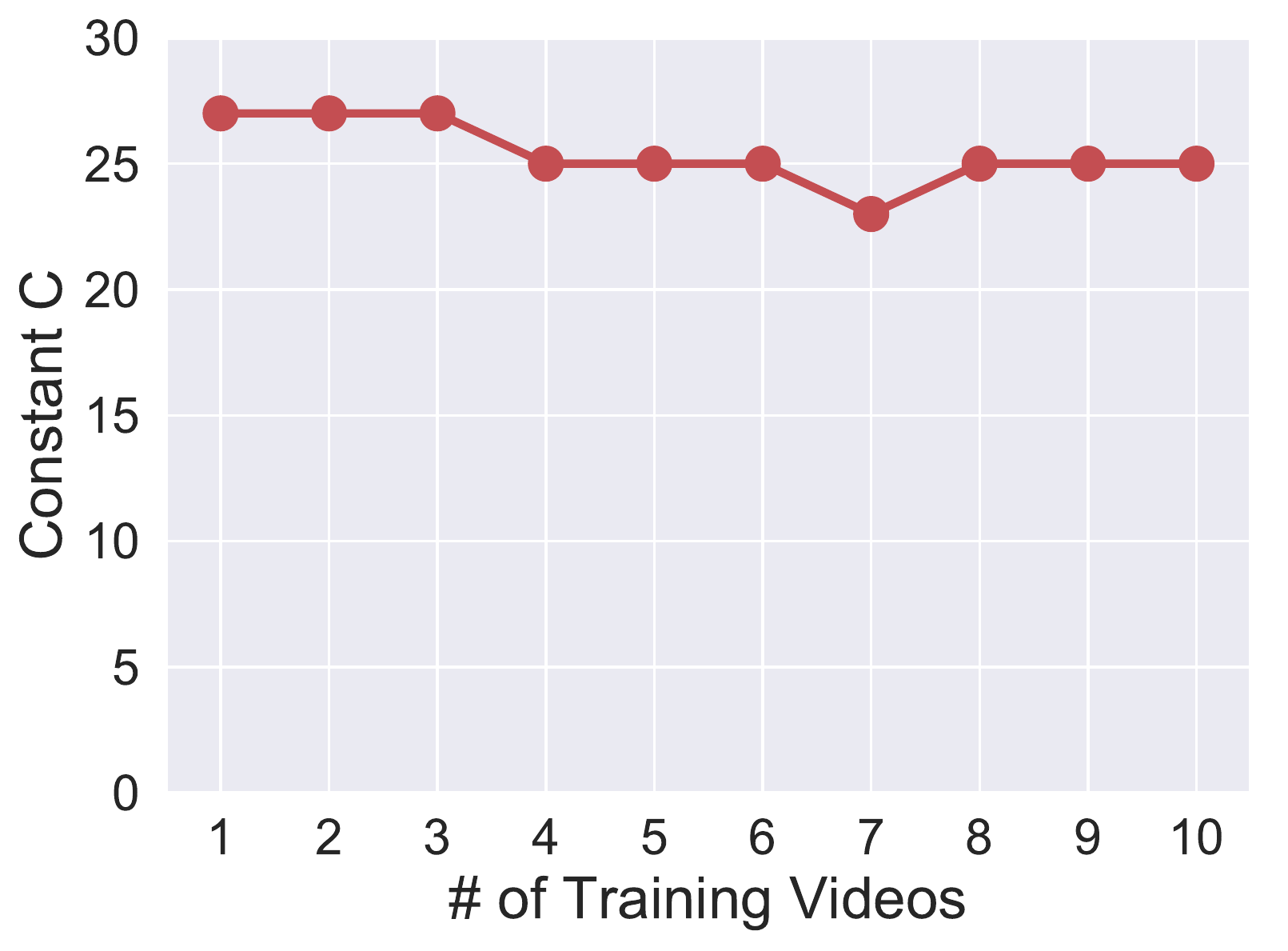}\vspace{-0.5em}
                \caption{Effect of Training Size}
                \label{fig:const-c}
        \end{subfigure}
        \caption{Evaluation of Adjustment Stage.}
        \label{fig:initializer-adjust}\vspace{-.5em}
\end{figure}

\vspace{.25em}

{\noindent \bf Adjustment Stage.} Suppose the prediction stage returns $k$ sliding windows as highlights. Then, the adjustment stage aims to find the approximate start positions of the highlights (i.e., red dots). It first finds the peak in each sliding window and subtracts it by a constant value $c$  (learned from labeled data) to get the red dot. We used ten videos as training data to get the constant value, and evaluated \videoprecisionstartK on fifty testing videos. The ideal situation of the adjustment stage is to be able to get a correct red dot for every top-$k$ highlight. So the Ideal curve in Figure~\ref{fig:adjust} is the same as the red line in Figure~\ref{fig:predict}. \toretter is a social-network based highlight detection approach. It detects highlights based on the time-series curve of message number. We compare with it to examine the necessity of having the adjustment stage in Highlight Initializer. From Figure~\ref{fig:adjust}, we can see that \toretter's precision was below $20\%$. Our adjustment method outperformed \toretter by around $3\times$.
This result validates that i) there is indeed a delay between the start of a highlight and the discussion of the highlight; ii) our adjustment method can capture the delay well.


We investigate how robust the constant value $c$ is by varying the number of training videos.
Figure~\ref{fig:const-c} shows the result. We can see that the constant value kept stable. It was in the range from 23s to 27s. This is because that users tend to have similar behaviors when watching highlights. It could be considered as "reaction time" of viewers, so that we can use a small amount of training data to generate a quite accurate constant value for adjustment.

\subsection{Evaluation of Highlight Extractor}\label{subsec:extractor-exp}

\begin{figure}[t]\vspace{-2.5em}
        \begin{subfigure}[b]{0.235\textwidth}
                \includegraphics[width=\linewidth]{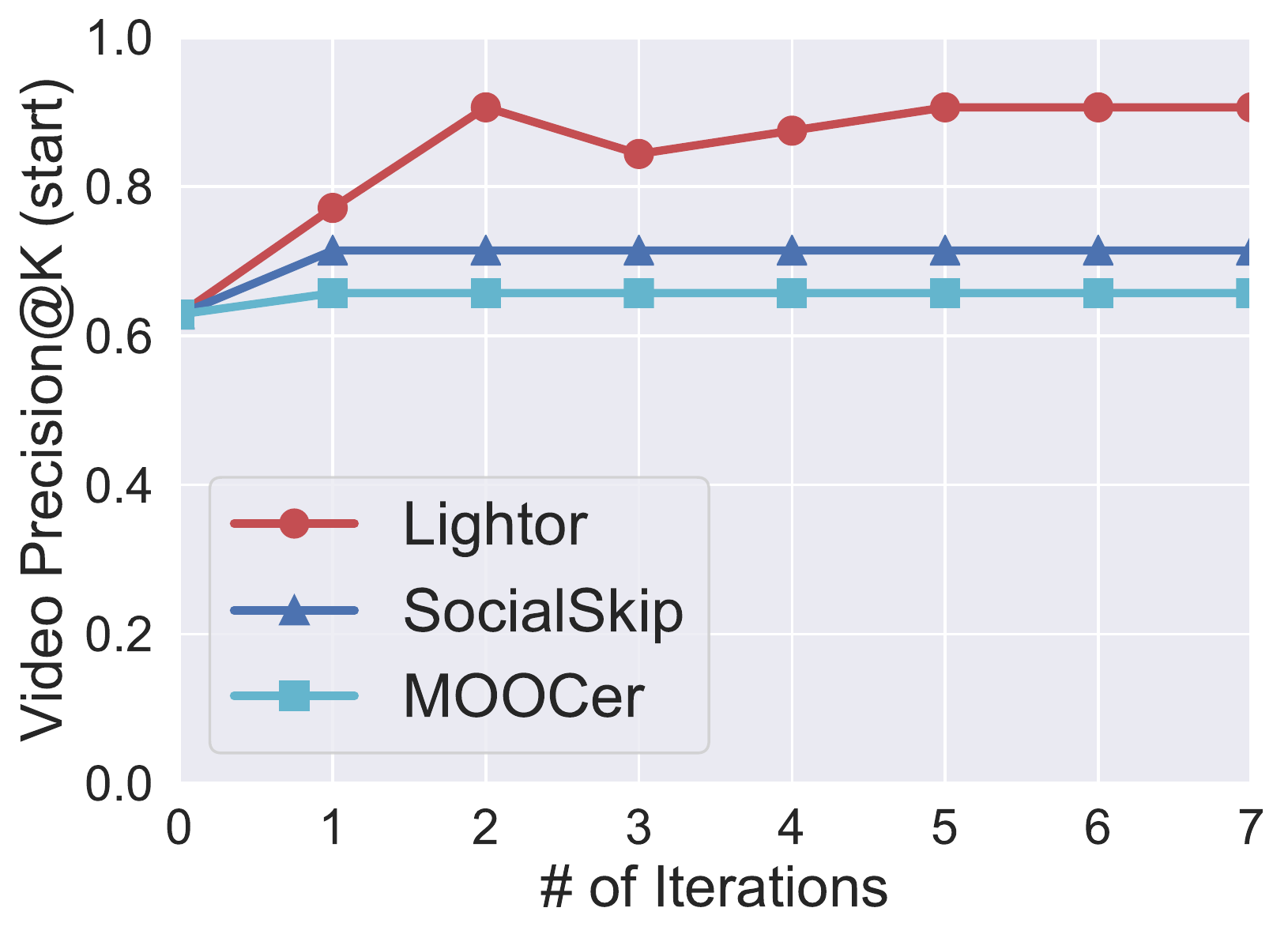}\vspace{-0.5em}
                \caption{Start Position}
                \label{fig:start}
        \end{subfigure}
        \begin{subfigure}[b]{0.235\textwidth}
                \includegraphics[width=\linewidth]{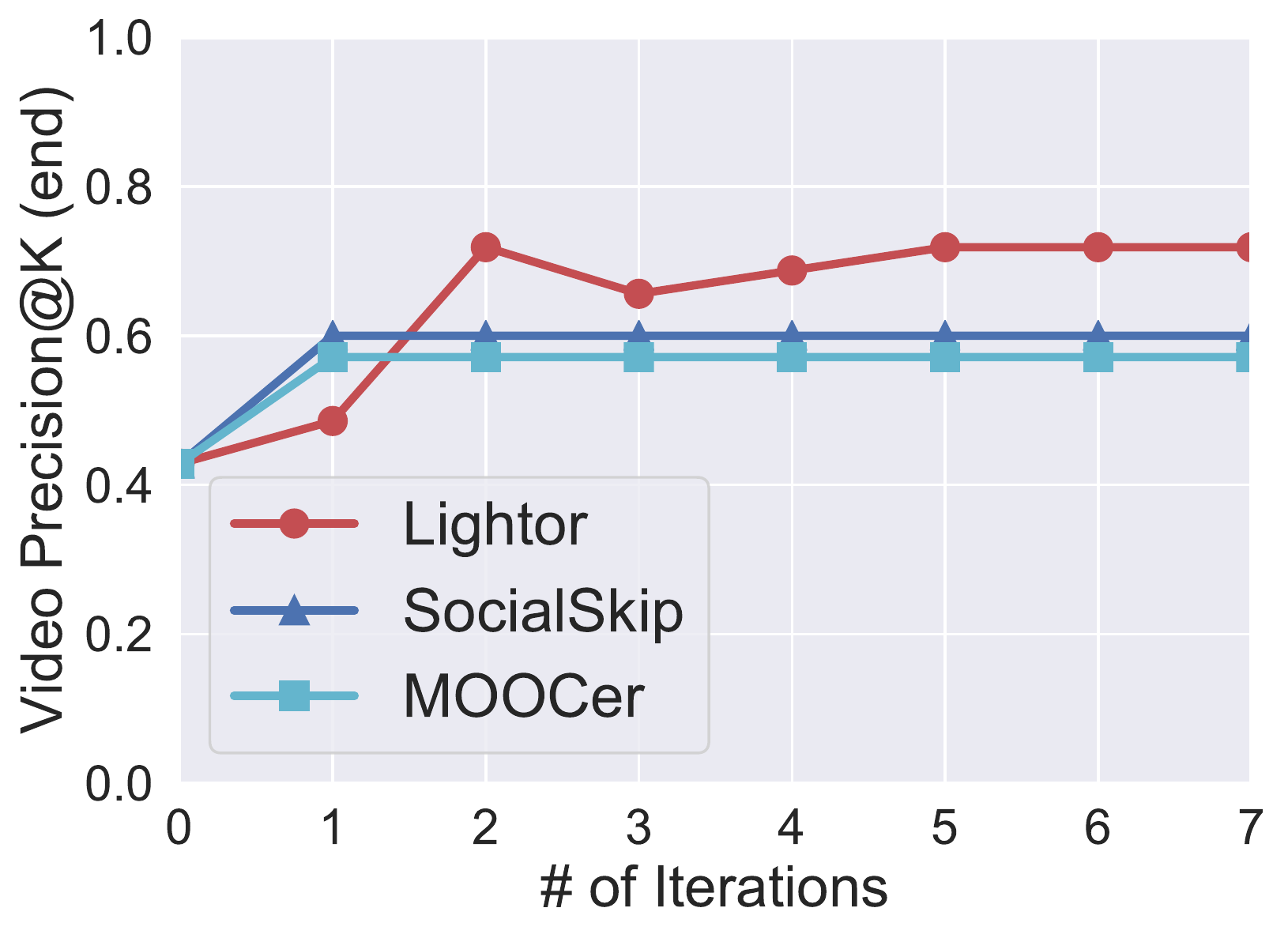}\vspace{-0.5em}
                \caption{End Position}
                \label{fig:end}
        \end{subfigure}
        \caption{Evaluation of Highlight Extractor.}
        \label{fig:extractor}\vspace{-1em}
\end{figure}

Highlight Extractor aims to leverage play data to identify the boundary of each highlight.  We compared our method with \socialskip\cite{Chorianopoulos2013} and \moocer\cite{Kim:2014:UID:2556325.2566237}. Both methods model the interaction data as a histogram along the timeline. Each bin represents a 1-second video clip, and the bin height represents how interesting the video clip is.  \socialskip leverages Seek Backward \& Forward to construct the histogram. When Seek Forward/Backward happens, it indicates that the range that the user jumps is interesting/uninteresting. The bin height would be added by +1/-1 accordingly. Once all user interactions are collected, \socialskip smooths the histogram curve and finds all local maxima of the curve. \socialskip subtracts each local maximum by 10s as the start position and adds it by 10s as the end position of a highlight. Differently, \moocer only leverages Play interactions. When Plays happen, the histogram range being played will be added by 1. After smoothing, \moocer finds the local maxima as well. For each local maximum, it finds two turning points aside the local maximum as the start and end position of a highlight.

We randomly selected 7 testing videos, and applied Highlight Initializer on them to generate 35 red dots (5 per video). We created  one task for each red dot.  We first published the 35 tasks to AMT. After receiving 10 responses for each task, we computed the new position of each red dot, and published a set of new tasks with updated red-dot positions to AMT. We repeated this process until users reached a consensus on the extracted highlights. Since \socialskip and \moocer  are not iterative, we applied them using our first iteration of interaction data. Figure~\ref{fig:extractor} shows how \videoprecisionstartK  and \videoprecisionendK change over iterations.  We can see that \proj kept improving over iterations, and outperformed \socialskip and \moocer by a big margin in the last iteration. This improvement came from two sources. On one hand, it removed the red dots that did not talk about a highlight (i.e., improving the prediction stage in Highlight Initializer); on the other hand, it made a better adjustment about where a red dot should be put (i.e., improving the results of adjustment stage in Highlight Initializer). 

\subsection{Applicability of \textit{\proj} in Twitch}

Based on our experiments, in order to achieve high precision, Highlight Initializer requires the number of chat messages per hour larger than 500 and Highlight Extractor requires more than 100 viewers per video. We examine the applicability of \proj with these requirements in Twitch. 

\begin{figure}[t] \vspace{-2.5em}
        \begin{subfigure}[b]{0.236\textwidth}
                \includegraphics[width=\linewidth]{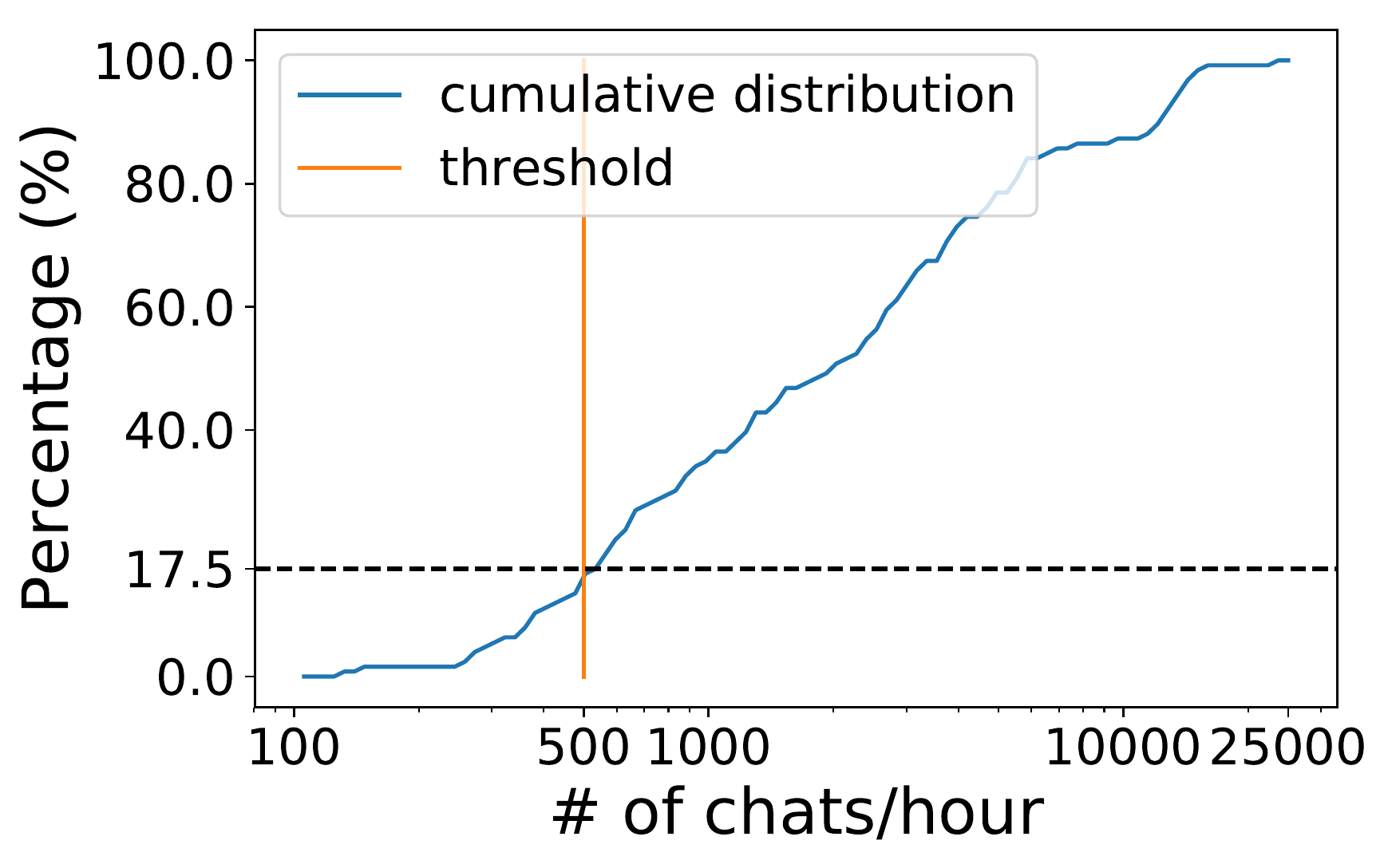}\vspace{-0.5em}
                \caption{CDF for Chat Messages}
                \label{fig:chat_cdf}
        \end{subfigure}
        \begin{subfigure}[b]{0.236\textwidth}
                \includegraphics[width=\linewidth]{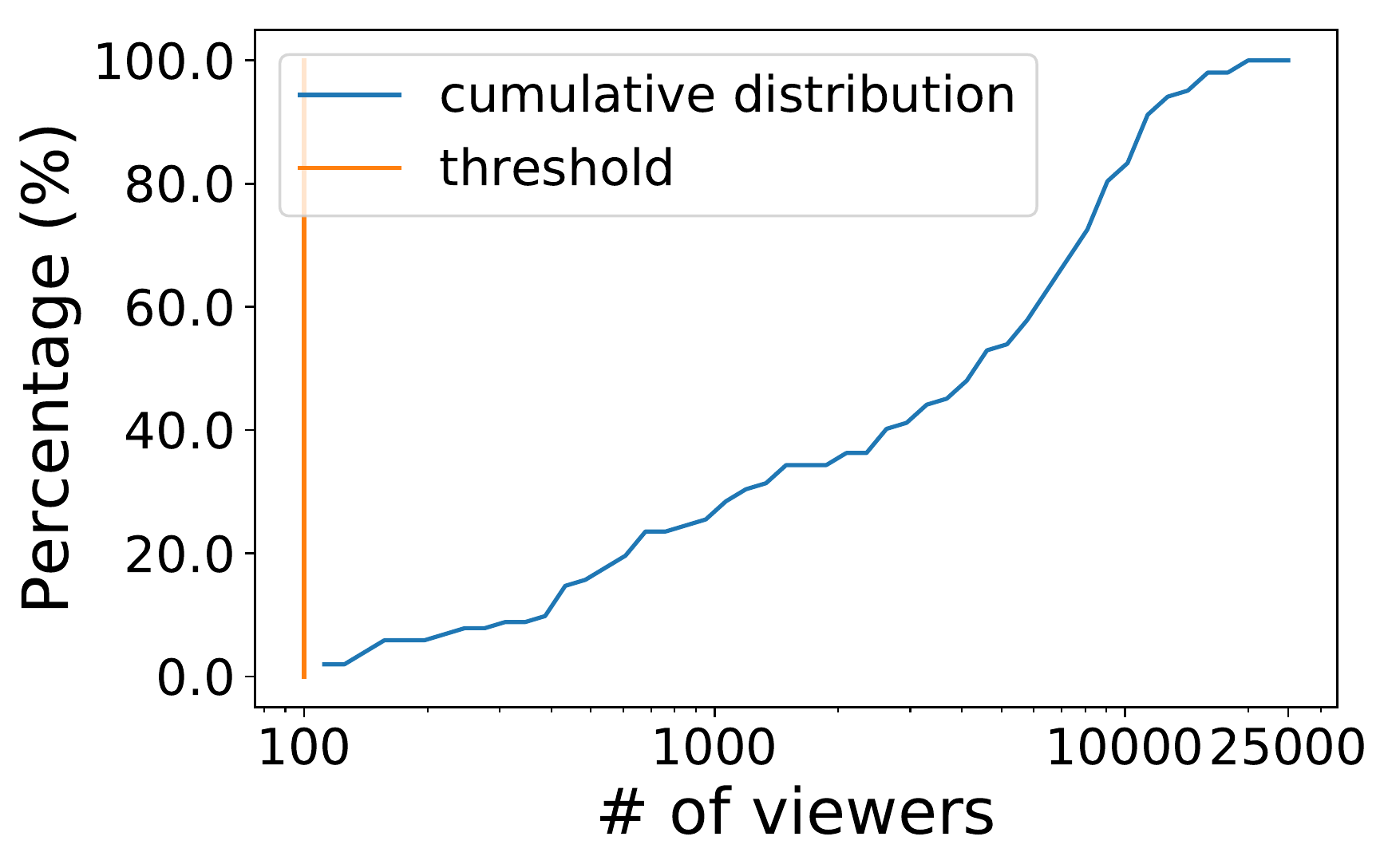}\vspace{-0.5em}
                \caption{CDF for Viewers}
                \label{fig:viewer_cdf}
        \end{subfigure}
        \caption{Cumulative distribution over recorded videos.}
        \label{fig:cdf} \vspace{-1em}
\end{figure}

We selected the top-10 channels in Dota 2 and crawled twenty most recently recorded videos from each channel. We plot the CDF of the number of chat messages and the number of viewers, respectively. Figure~\ref{fig:cdf} shows the results.  We see that more than $80\%$ of recorded videos have more than 500 chat messages per hour and all the recorded videos have more than 100 viewers. These results indicate that \proj is applicable to the majority of  popular videos  in Twitch.

\subsection{Comparison with Deep Learning}\label{subsec:deeplearning}

We compared \proj with the state-of-the-art deep learning approach~\cite{DBLP:conf/emnlp/FuLBB17}. We first compared \proj's Highlight Initializer with \chatlstm since both of them use chat messages only. Then we conducted an end-to-end comparison between \proj and \jointlstm.

\chatlstm is a character-level 3-layer LSTM-RNN~\cite{DBLP:journals/corr/Graves13} model. For each labeled frame, it treats all chat messages that occur in the next 7-second sliding window as input. \jointlstm is built on top of a video model and \chatlstm. The video model uses a memory-based {\small LSTM-RNN} on top of image features extracted from pre-trained image models. We used \lol dataset to train models and applied on both \lol and \dota datasets. We applied \chatlstm and \jointlstm to predict the probability of each frame being a highlight, and selected the top-k frames. Close frames usually mean they belong to the same highlight, thus if two frames are close to each other (within 120s which is consistent with our setting in Section \ref{subsec:obj}), we only pick up the frame with a higher probability.

\sloppy

\vspace{.5em}

\begin{figure}[t]\vspace{-1.5em}
        \begin{subfigure}[t]{0.235\textwidth}
                \includegraphics[width=\linewidth]{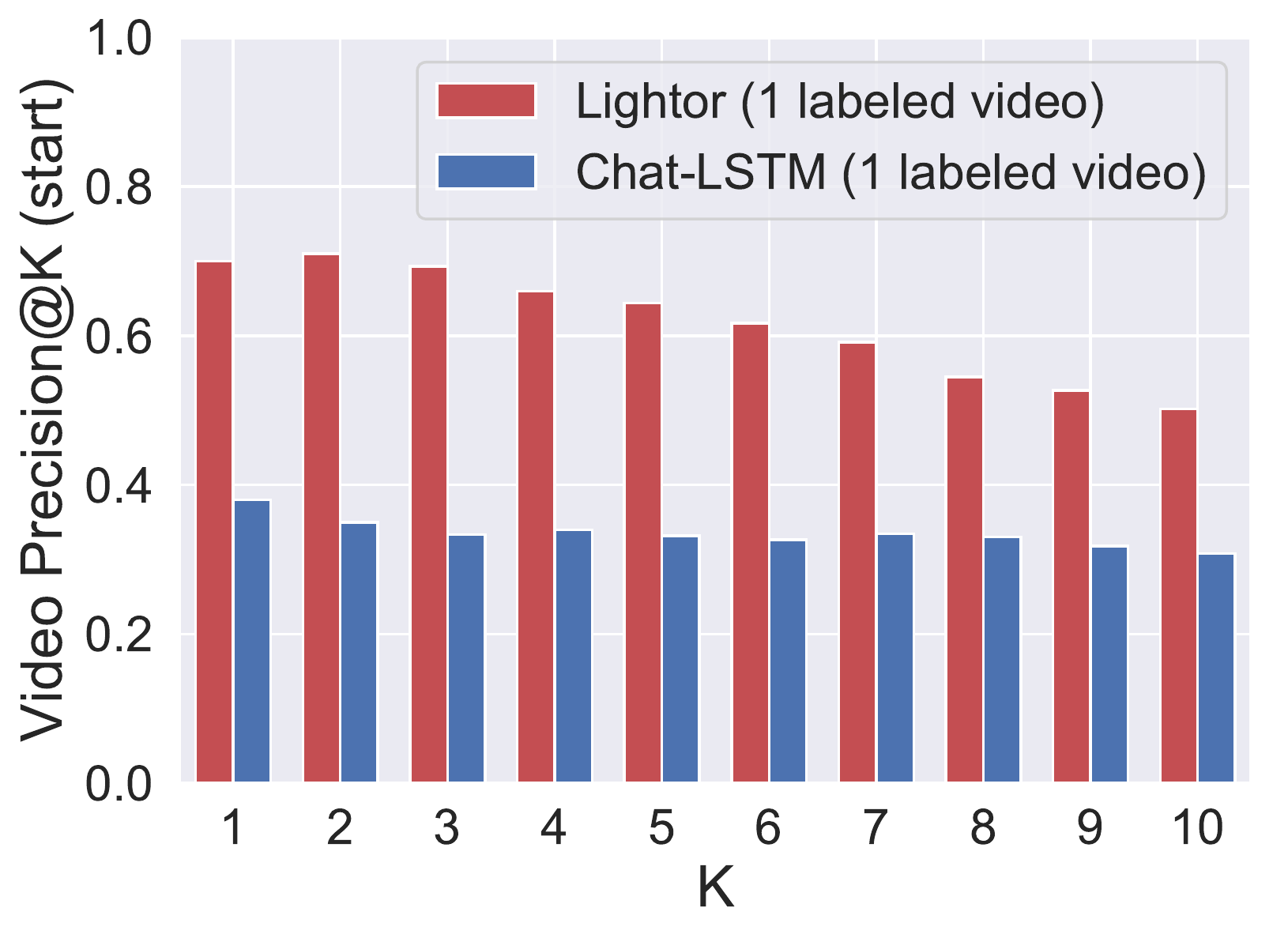}\vspace{-0.5em}
                \caption{ Both \proj and \scriptsize{\textsf{Chat-LSTM}} were trained on 1 video.}
                \label{fig:dl-1-label}
        \end{subfigure}\hspace*{0.5em}
        \begin{subfigure}[t]{0.235\textwidth}
                \includegraphics[width=\linewidth]{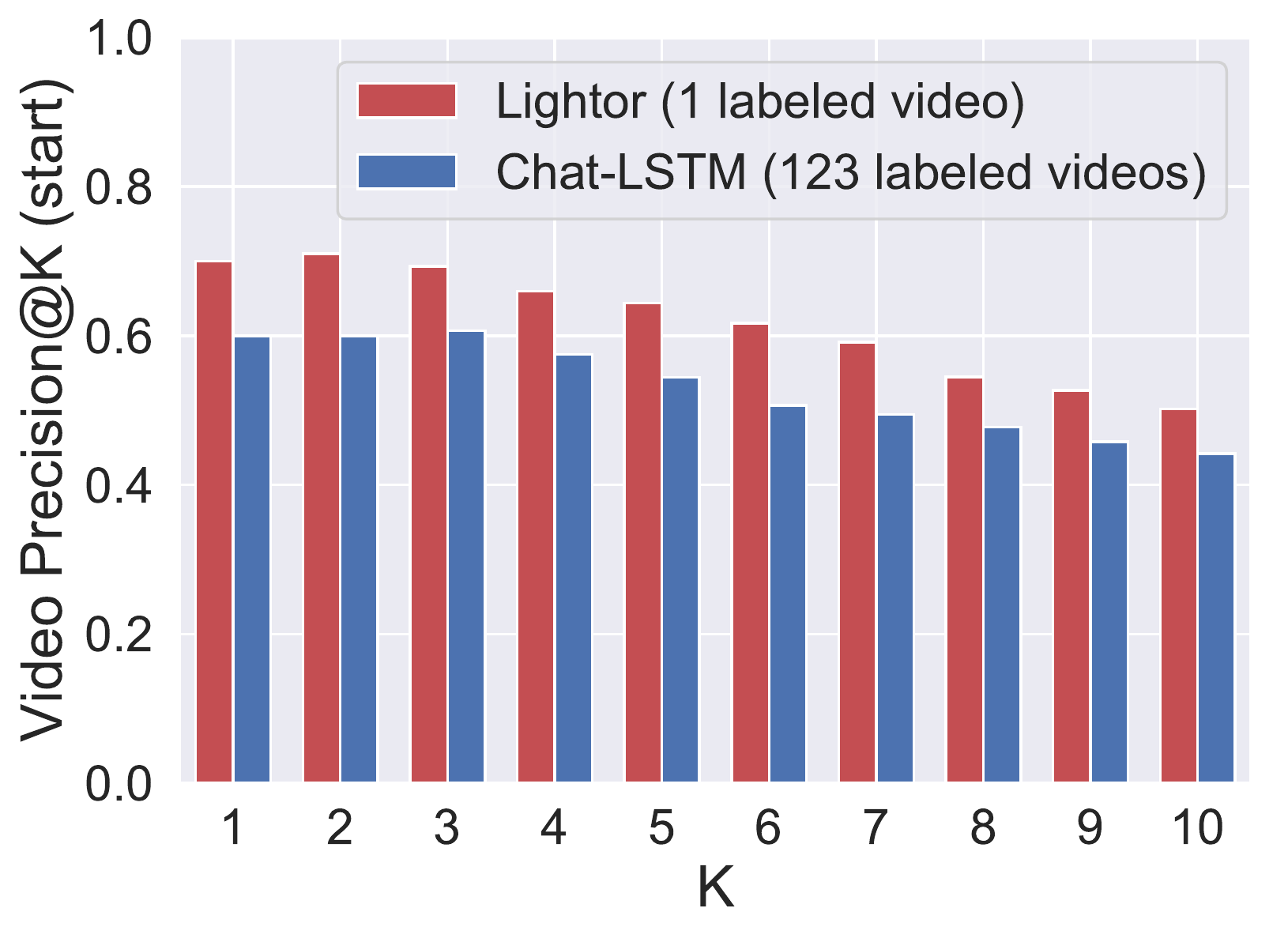}\vspace{-0.5em}
                \caption{\proj was trained on 1 video while \scriptsize{\textsf{Chat-LSTM}} was trained on 123 videos.}
                \label{fig:dl-123-label}
        \end{subfigure}
        \caption{Comparison of \proj and \scriptsize{\textsf{Chat-LSTM}} in terms of training data size (LoL data, 50 test videos).}
        \label{fig:cmp-trainsize}\vspace{-.5em}
\end{figure}

{\noindent \bf Comparison with \chatlstm.} We first compare \proj with \chatlstm in terms of training data size. Figure~\ref{fig:dl-1-label} shows the result. We can see that \proj only needed to label a single video in order to achieve a high precision, but \chatlstm did not perform well with a single labeled video. This is because that \proj detects highlights based on a small number of generic features. We increased the training size of \chatlstm to 123 labeled videos, and compared with \proj with 1 labeled training video. As shown in Figure~\ref{fig:dl-123-label}, \chatlstm's performance got improved but still performed worse than \proj. This is because that \chatlstm is not good at adjusting delay between chat messages and video contents. 

\begin{figure}[t]\vspace{-1.5em}
        \begin{subfigure}[b]{0.235\textwidth}
                \includegraphics[width=\linewidth]{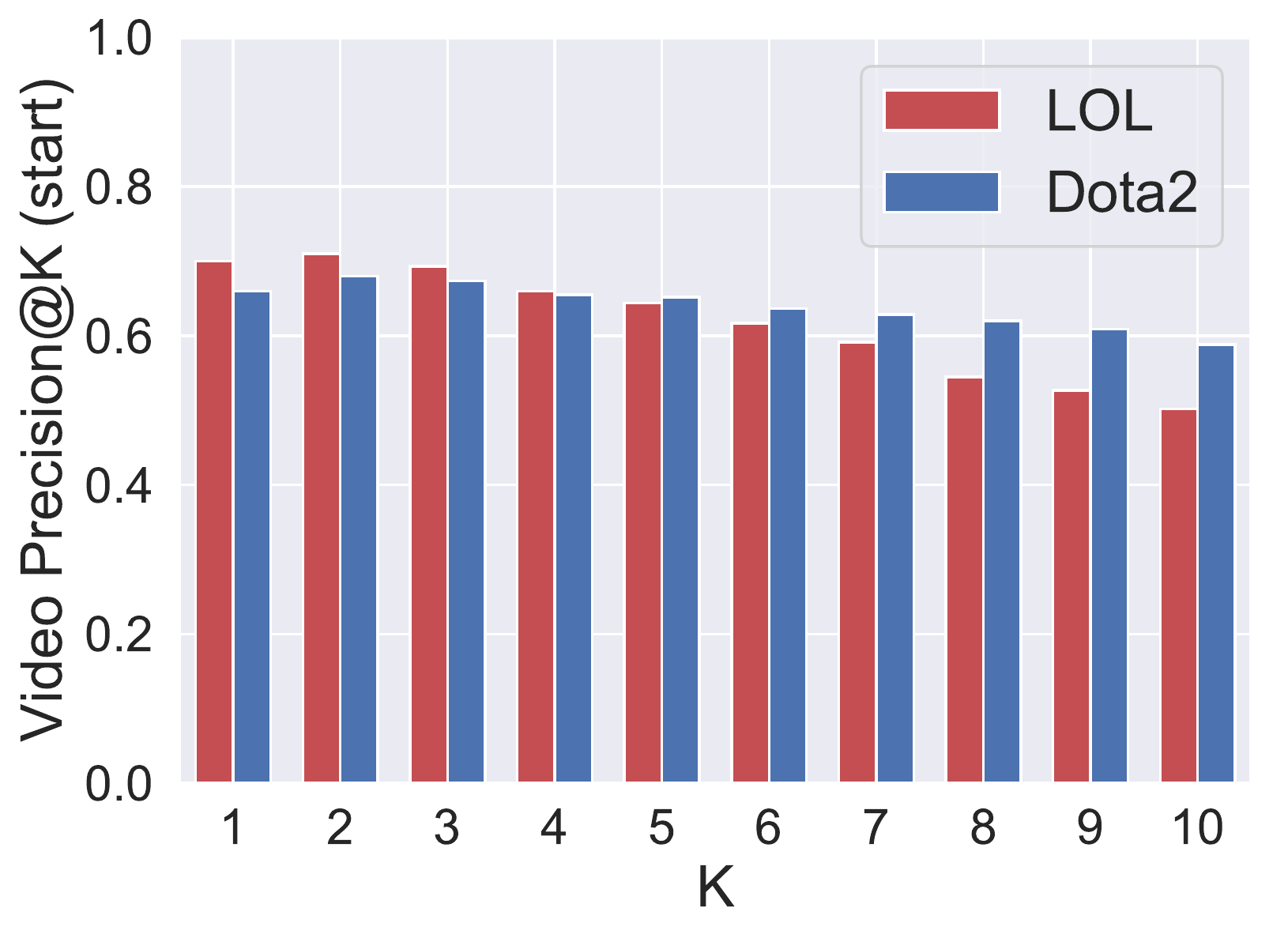}\vspace{-0.5em}
                \caption{\proj was trained on LoL, and tested on LoL and Dota 2}
                \label{fig:lightor-generalization}
        \end{subfigure}\hspace*{0.5em}
        \begin{subfigure}[b]{0.235\textwidth}
                \includegraphics[width=\linewidth]{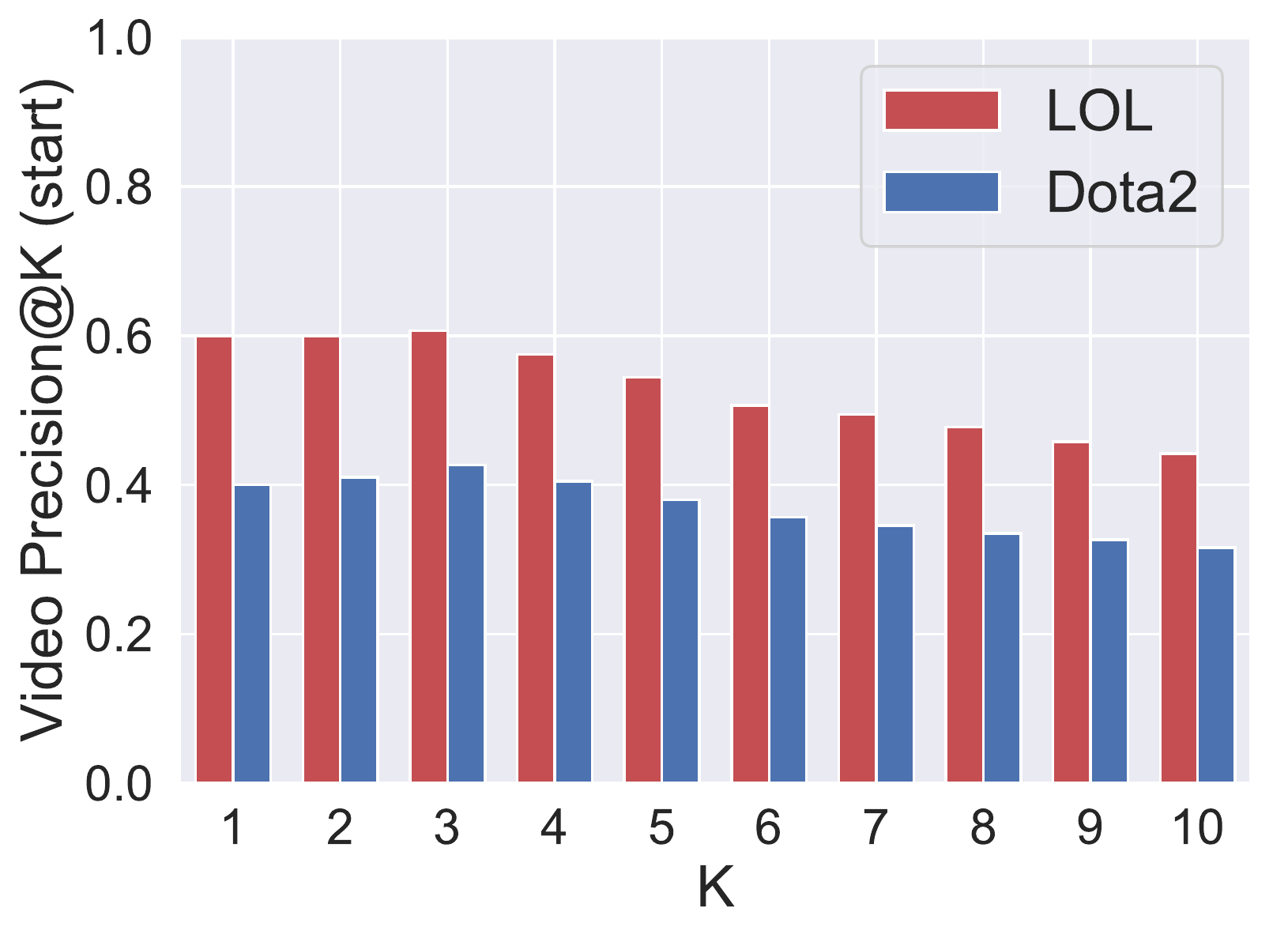}\vspace{-0.5em}
                \caption{\scriptsize{\textsf{Chat-LSTM}} was trained on LoL, and tested on LoL and Dota 2}
                \label{fig:dl-generalization}
        \end{subfigure}
        \caption{Comparison of \proj and \scriptsize{\textsf{Chat-LSTM}} in terms of model generalization.}
        \label{fig:cmp-generalization}\vspace{-.5em}
\end{figure}

We then compare \proj with \chatlstm in terms of model generalization. Figure~\ref{fig:cmp-generalization} shows the result. In Figure~\ref{fig:lightor-generalization} we can see that for \proj, the model trained on one game type (\lol) can still achieve high precision on another game type (\dota).  \proj has good generalization because the selected features are very general. Interestingly, \proj performed even better on \dota data for $k > 5$. This is because \dota videos have more highlights. For \chatlstm, as shown in Figure~\ref{fig:dl-generalization}, there is a big performance gap between \lol and \dota. That is, if we trained the \chatlstm model on one game type (\lol) and then tested it on another game type (\dota), the model did not generalize well.  

\vspace{.5em}

{\noindent \bf Comparing with \jointlstm.} We compare \proj with \jointlstm on their end-to-end performance. \proj was trained on 1 labeled \lol video and collected user interactions, while \jointlstm was trained on 123 labeled \lol videos. We tested them on seven \dota videos. Table~\ref{table:end-to-end} reports their training time and Video Precision of Top-5 highlights (k=5). In terms of efficiency, \proj required 100000$\times$ less training time compared to \jointlstm. In terms of effectiveness, \proj achieved a \videoprecisionstartK of 0.906 and a \videoprecisionendK of 0.719, while the \videoprecisionstartK and the \videoprecisionendK of \jointlstm are both round 0.6. This is because that \proj has a much better generalization than \jointlstm. 

\begin{table}[h] 
\centering
\begin{tabular}{|l | c  c   c|} 
 \hline
 \multirow{2}{*}{\textbf{Systems}}  & \textbf{Precision@K} & \textbf{Precision@K} & \textbf{Training time}\\
  & \textbf{(Start)} & \textbf{(End)} & \\
 \hline
 \proj & 0.906 & 0.719 & 1.06 sec\\ 
  \hline
 \footnotesize{\textsf{Joint-LSTM}} &  0.629 & 0.600 & $>$3 days\\
 \hline
\end{tabular}
\caption{\small An end-to-end comparison between \proj and \jointlstm. }
\label{table:end-to-end} \vspace{-.5em}
\end{table}

In summary, the experimental results indicate that \proj has great advantages over these deep-learning based approaches in terms of training data size, computational cost and generalization. Nevertheless, we do not argue to totally replace the deep-learning based approach with \proj. Deep learning has its own advantages. For example, if a deep learning model is trained over video data, it does not need chat messages or user interaction data to detect highlights. An interesting future direction is to explore how to combine \proj with Deep Learning, where \proj is used to generate high-quality labeled data and Deep Learning is then applied to train a model.

\section{Findings \& Lessons learned} \label{sec:lessons}
We present interesting findings and lessons learned. 

\sloppy
\begin{itemize}[leftmargin=*]\setlength\itemsep{.15em}
    \item It is important to do a pilot test and analyze real user interaction data. For example, we originally thought that Seek Backward could be a useful indicator to detect start positions of highlights. However, from real data we find that since there are various reasons to trigger this interaction (e.g., re-watch a highlight, look for a new highlight), it is not easy to infer users' true intent.
    \item The recorded live videos in Twitch typically attract thousands of viewers on average. In our experiments, we recruited around 500 viewers and showed promising results. Based on these findings, we believe that there is no obstacle for \proj to collect enough user interaction data in real live streaming platforms.
    \item Users prefer spreading the red dots over the entire progress bar, instead of cluttering them in a narrow region. They think  the former can help them have a broader overview of the whole video, while the latter only shows the content of part of the video.
    \item Viewers sometimes get excited about the interesting clips that are not related to a video's main theme, such as the break between two games, or the preparation for a game. \proj may identify these clips as highlights. We will study how to overcome this limitation in future work.
\end{itemize}




\section{Conclusion \& Future work} \label{sec:conc}

\sloppy 

We presented \proj, a novel implicit crowdsourcing workflow to extract  highlights for recorded live videos. \proj consists of two components. In Highlight Initializer, we explored different design choices and justified our decisions. We proposed three generic features (message number, message length, and message similarity) and built a model to predict highlight positions. We also noticed that there is a delay between a highlight and its comments, and proposed a simple learning-based approach to estimate the delay. In Highlight Extractor, we identified the challenges to use noisy user interaction data to extract highlights, and proposed a three-stage dataflow (filtering $\rightarrow$ classification $\rightarrow$ aggregation) to address these challenges. We discussed how to implement \proj as a web browser extension and how to integrate \proj into existing live streaming platforms. We recruited about 500 real users and evaluated \proj using real Dota 2 and LoL data. We compared with various baselines in the social network, online learning, and deep learning fields.  The results showed that \proj achieved very high detection accuracy (\precisionK: 70\%-90\%). Furthermore, it only needed to label a single video and spend a few seconds on training, and the obtained model had good generalization.  

\fussy

There are many future research directions to explore. First, we were told by the data science team at a well-known live streaming platform that they stored several terabytes of chat data, but have not tried to extract value from the data. We are planning to deploy \proj on their platform, and conduct more large-scale real-world experiments. We would like to extensively test our system in terms of the volumes and diverse types of videos. Second, we want to further optimize the workflow, especially on the adjustment stage. The current implementation assumes that there is a simple linear relationship between $\timepeak$ and $\timestart$. We plan to relax this assumption and build a more sophisticated regression model. Third, we plan to further evaluate the generalization of our system using data collected from other domains (e.g., celebrity events) and other live streaming platforms (e.g., YouTube Live). Fourth, this paper demonstrates a great potential of the application of implicit crowdsourcing to video highlight detection. It is promising to investigate how to design an implicit crowdsourcing workflow for other video analysis tasks (e.g., video querying, video summarization, and video indexing).

\bibliographystyle{IEEEtran}
\bibliography{ref}

\end{document}